\documentclass[10pt]{article}
\usepackage[utf8]{inputenc}
\usepackage{cite}
\usepackage{hyperref}
\hypersetup{colorlinks=true,linkcolor=bluecyan,citecolor=orange3,urlcolor=green3,pdfencoding=auto,linktocpage}
\usepackage{cancel}
\usepackage{amsmath,amssymb,amsbsy,amstext,amsthm,simplewick,amsfonts,braket}
\usepackage{braket}
\usepackage{mathrsfs}
\usepackage{graphicx}
\usepackage{wrapfig}
\usepackage{upgreek}
\usepackage{bm} %-> bold math
\usepackage{framed}
\usepackage{bbm}
\usepackage{textcomp}
\usepackage{adjustbox}
\usepackage{makecell}
\usepackage{tcolorbox}
\usepackage{empheq}
\usepackage[normalem]{ulem}
\usepackage{enumitem}
\usepackage{array}
\usepackage{dsfont}
\usepackage{ulem}
\usepackage{xcolor}
\usepackage{caption}
\usepackage{subcaption}
\usepackage{simpler-wick}
\usepackage{tocloft}
\usepackage{tikz}
\usepackage[compat=1.1.0]{tikz-feynman}
\usepackage{relsize}
\usepackage{float}
\tikzstyle{arrow} = [thick,->,>=stealth]

\usepackage[framemethod=default]{mdframed}
\usetikzlibrary{shapes.geometric, arrows}

\tikzstyle{startstop} = [rectangle, rounded corners, 
minimum width=3cm, 
minimum height=1cm,
text centered, 
draw=black, 
fill=white!30]

\newmdenv[backgroundcolor=gray!15,%
skipabove=5pt,%
skipbelow=5pt,%
leftmargin=2pt,%
rightmargin=2pt,%
innertopmargin=-6pt,%
innerbottommargin=5pt,%
innerleftmargin=5pt,%
innerrightmargin=5pt,%
splittopskip=0pt,%
splitbottomskip=0pt,%
linewidth=0pt,%
nobreak=true]%
{keyeqn}

\newmdenv[backgroundcolor=gray!15,%
skipabove=5pt,%
skipbelow=5pt,%
leftmargin=2pt,%
rightmargin=2pt,%
innertopmargin=-2pt,%
innerbottommargin=5pt,%
innerleftmargin=5pt,%
innerrightmargin=5pt,%
splittopskip=0pt,%
splitbottomskip=0pt,%
linewidth=0pt,%
nobreak=true]%
{keythrm}

\graphicspath{{Fig/}}

\definecolor{lightgreen}{cmyk}{0.2, 0, 0.2, 0.2}
\definecolor{lightgray}{cmyk}{0.1,0.2,0,0.1}
\definecolor{lightgray2}{cmyk}{0.1,0.1,0,0.1}
\definecolor{bluecyan}{RGB}{0, 100, 200}
\definecolor{blue3}{RGB}{31,119,180}
\definecolor{red3}{RGB}{214,39,40}
\definecolor{orange3}{RGB}{255,127,14}
\definecolor{green3}{RGB}{44,160,44}

\definecolor{red2}{RGB}{255,0,0}
\definecolor{green2}{RGB}{0,170,0}
\definecolor{blue2}{RGB}{0,128,255}
\definecolor{magenta2}{RGB}{191,64,191}
\definecolor{purple2}{RGB}{112,48,160}
\definecolor{orange2}{RGB}{255,192,0}

\newcommand\Ccancel[2][red]{
    \let\OldcancelColor\CancelColor
    \renewcommand\CancelColor{\color{#1}}
    \cancel{#2}
    \renewcommand\CancelColor{\OldcancelColor}
}

\numberwithin{equation}{section}

\allowdisplaybreaks[1]
\begin{document}

\begin{titlepage}
	\setcounter{page}{1} \baselineskip=15.5 pt 
	\thispagestyle{empty}

 \begin{center}
		{\fontsize{18}{18}\centering {\bf{Hidden Adler zeros and soft theorems for \\ \vspace{0.2cm} inflationary perturbations}}\;}\\
	\end{center}
 
	\vskip 18pt
	\begin{center}
		\noindent
		{\fontsize{12}{18}\selectfont Zong-Zhe Du\footnote{\tt zongzhe.du@nottingham.ac.uk}$^{,a}$}
	\end{center}
	
	\begin{center}
		\vskip 8pt
		$a$ \textit{School of Physics and Astronomy,
			University of Nottingham, University Park, \\ Nottingham, NG7 2RD, UK} 
	\end{center}
	
	%=========================================

	\noindent \textbf{Abstract}  We derive soft theorems for on-shell scattering amplitudes from non-linearly realised global space-time symmetries, arising from the flat space and decoupling limits of the effective field theories (EFTs) of inflation, while taking particular care of on-shell limits, soft limits, time-ordered correlations, momentum derivatives, energy-momentum conserving delta functions and $i\varepsilon$ prescriptions. Intriguingly, contrary to common belief, we find with a preferred soft hierarchy among the soft momentum $q$, on-shell residue $p_a^0 \pm E_a$, and $\varepsilon$, the soft theorems do not have dependence on unconstrained off-shell interactions, even in the presence of cubic vertices. We also argue that the soft hierarchy is a natural choice, ensuring the soft limit and on-shell limit commute. Our soft theorems depend solely on on-shell data and hold to all orders in perturbation theory. We present various examples including polynomial shift symmetries, non-linear realisation of Lorentz boosts and dilatations on how the soft theorems work. We find that the collection of exchange diagrams whose soft momenta are associated with cubic vertices, that are indeterminate in the soft limit, exhibits an enhanced soft scaling. The enhanced soft scaling explains why the sum of such diagrams do not enter the soft theorems non-trivially. We further apply the soft theorems to bootstrap the scattering amplitudes of the superfluid and scaling superfluid EFTs, finding agreement with the Hamiltonian analysis.
	
	%=========================================
	
\end{titlepage} 

%\restoregeometry

\newpage
\setcounter{page}{2}
{
	\tableofcontents
}

%=======================================
%=======================================

\section{Introduction}

Soft limits of scattering amplitudes play crucial roles in identifying IR structures of effective field theories (EFTs) \cite{Cheung:2014dqa}. Scattering processes performed at the ground state that spontaneously breaks the original symmetry often exhibit universal features, e.g. the amplitude vanishes upon sending one external momentum to zero, i.e. $\mathcal{A}(p) \sim \mathcal{O}(p)$. This is known as the Adler zero \cite{Adler:1965ga}. When a space-time symmetry is spontaneously broken, we get additional constraints from the non-linear symmetries, and the scattering amplitude can enjoy an enhanced soft scaling $\mathcal{A}\sim \mathcal{O}(p^\sigma)$ where $\sigma$ is dictated by the specific non-linear realisation of the symmetry \cite{Cheung:2014dqa,Cheung:2016drk}. For example, in scalar Dirac-Born-Infeld (DBI) theory the $D+1$ dimensional Lorentz boost is non-linearly realised on the $D$ dimensional Minkowski space where $\phi$ is realised as the $D+1$ th coordinate. The upshot of the non-linear symmetry is that at tree level, there are surprising cancellations among different topologies such that the amplitude scales as $\mathcal{A}(p) \sim \mathcal{O}(p^2)$ despite the Lagrangian having only one derivative per field. Another example is the special Galileon which enjoys an enhanced Adler zero $\mathcal{A}\sim \mathcal{O}(p^3)$ due to a hidden symmetry \cite{Hinterbichler:2015pqa}.
See \cite{Padilla:2016mno,Cheung:2015ota,Cheung:2018oki,Bonifacio:2019rpv,Brauner:2022ymm,CarrilloGonzalez:2019fzc,Bartsch:2022pyi,Goon:2020myi} for more work on single scalar soft theorems when Poincar\'{e} symmetry is linearly realised.

One crucial property of Lorentz invariant single scalar field EFTs is that shift symmetry does not allow non-trivial cubic vertices, namely the 3-point amplitude would always vanish even after analytical continuation. This implies that cubic vertices can be removed by a field redefinition. There are two ways to endow shift symmetric EFTs with non-vanishing 3-point amplitudes, one is to couple them to more degrees of freedom and the other is to break Lorentz symmetries. The former scenario has been widely explored in recent years, including scattering amplitudes for non-linear sigma models and their hidden relations with the Tr$\phi^3$ theories. The effect of coloured structures on soft theorems and zeros were studied in \cite{Cheung:2021yog,DiVecchia:2015jaq,Derda:2024jvo,Kampf:2019mcd,Arkani-Hamed:2023swr, Li:2024qfp}.  The latter scenario is inextricably related to inflationary cosmology. Such boost-breaking theories can arise from taking the decoupling and flat-space limit of EFT of inflation \cite{Cheung:2007st}. The corresponding scattering amplitude, computed by taking such limits of the theory, appears as the residue of the inflationary wavefunction coefficient in the limit where the sum of the magnitude of the external spatial momenta approaches zero \cite{Maldacena:2011nz}
\begin{align}
\lim_{k_{T} \rightarrow 0} \psi_{n} = (\text{normalisation factors}) \times \frac{\mathcal{A}_{n}}{k_{T}^{p}} \,,
\end{align}
where $\psi_n$ is the order $n$ inflationary wavefunction coefficient, $\mathcal{A}_{n}$ is a flat-space boost-breaking scattering amplitude, and $k_{a}$ represents magnitude of momenta with $k_{T} = \sum_{a=1}^{n} k_{a}$ being the ``total energy". The connection between cosmology and flat-space scattering amplitude is extensively discussed in \cite{Baumann:2022jpr, Benincasa:2022gtd, Benincasa:2022omn}, using scattering amplitude as input data for cosmological bootstrap is implemented in \cite{Bonifacio:2022vwa, Baumann:2020dch, Baumann:2021fxj, Jazayeri:2022kjy, Mei:2024abu}, soft theorems for Lorentz breaking theories were studied in \cite{Green:2022slj,Mojahed:2022nrn,Du:2024hol,Cheung:2023qwn}. In both scenarios, it is claimed Adler zeros do not hold in general even at tree level since the propagator attached to a 3-point vertex would generate a soft pole even in the presence of a shift symmetry. As a consequence, the 3-point functions are either assumed to not be present by construction or explicitly subtracted such that the rest of the amplitude satisfies the soft theorem \cite{Cheung:2021yog,DiVecchia:2015jaq,Derda:2024jvo,Kampf:2019mcd,Du:2024hol,Brauner:2022ymm ,Green:2022slj, Cheung:2023qwn}. However, it was shown in \cite{Hui:2022dnm}, that the soft theorem for correlators are insensitive to exchange diagrams containing soft cubic vertices. This raises the question whether we could come up with soft theorems for scattering amplitudes that are also independent of cubic vertices that cannot be removed by field redefinitions. Another problem is understanding the meaning of taking derivatives with respect to soft momentum.  For Lorentz invariant theories, the derivatives are manifest, since both the interaction and on-shell condition preserve Lorentz invariance, the theory is free from any basis we choose. For boost breaking theories, the situation is different as can be been seen from an example with a cubic interaction $\dot\pi^3$. The 3-point amplitude is
\begin{align}
    \mathcal{A}_{3} \sim p_1^0 p_2^0 p_3^0.
\end{align}
The naive soft theorem arising from the symmetry transformation $\delta\pi = b_i x^i$  would be
\begin{align}
    \lim_{\vec{p}\rightarrow 0} \partial_{\vec{p}}\mathcal{A} = 0,
\end{align}
and even if we choose the basis where we eliminate energy by momentum dependence the amplitude does not satisfy this soft theorem despite it coming from an invariant interaction. In \cite{Du:2024hol} this ambiguity is overcome by taking particular care of the order of soft limit, on-shell limit and momentum derivative while not neglecting the time integral from $-\infty$ even in the positive energy branch where the structure of amplitude combination $\tilde{\mathcal{A}}_{E_{p}} \pm \tilde{\mathcal{A}}_{-E_{p}}$ emerges. The new soft theorem with the combination structure highly resembles the wavefunction coefficient soft theorem derived in \cite{Bittermann:2022nfh}. It is also shown in \cite{Du:2024hol} that the tower structure derived from acting with momentum derivatives on momentum conserving delta functions precisely reflects what is derived from the coset construction and the inverse Higgs constraints. 

\begin{figure}[H]
\centering
\begin{tikzpicture}[node distance=3cm]
\node (3) [startstop] {Symmetry};
\node (2) [startstop, below of=3] {Ward-Takahashi identity};
\node (1) [startstop, below of=2] {Scattering amplitude};
\draw [arrow] (3) -- node [anchor = east]{Noether current $J^\mu$} (2);
\draw [arrow] (2) -- node [anchor = east] {LSZ reduction} (1);
\end{tikzpicture}
    \caption{Schematic route for deriving scattering amplitude constraints from symmetries}
    \label{fig:route}
\end{figure}
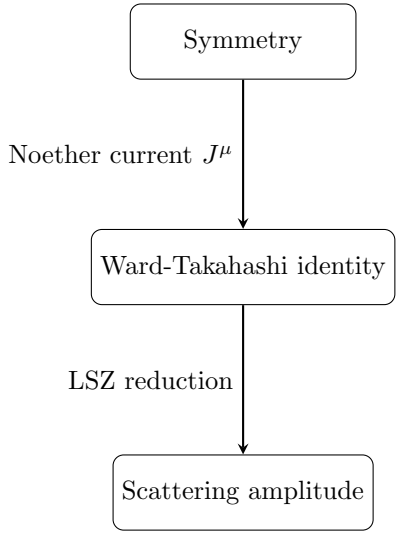

In the previous work \cite{Du:2024hol}, we restrict ourselves to field independent symmetries, now we introduce field-dependent terms within the symmetry transformation that could be applied to more realistic set ups. Our primary focus will be on deriving a soft theorem for the non-linear realisation of Lorentz boosts i.e. in a theory of fluctuations about a background $\braket{\phi} = t$. In such a theory the non-linear boost transformation for the Nambu-Goldstone mode is\footnote{The Lorentz boost symmetry $\delta\phi = b_i (x^i\partial_t + t\partial_i)\phi$ is non-linearly realised on the fluctuation around the time dependent VEV $\phi = t + \pi$ by 
\begin{align}
    \delta \pi = b_i [x^i + (x^i\partial_t + t\partial_i)\pi].
\end{align}
where the non-linear term $b_i x^i$ comes from boost operator acting on the VEV.}
\begin{align}
    \delta \pi = b_i [x^i + (x^i\partial_t + t\partial_i)\pi].
\end{align}
The amplitude soft theorem for this non-linear symmetry is derived in \cite{Green:2022slj,Cheung:2023qwn} where the form depends on explicit expressions of cubic vertices. In this paper we derive a new soft theorem that doesn't have such a dependence on cubic vertices. To derive constraints on on-shell amplitudes from symmetries, we begin by calculating the Noether current associated with the symmetry. This current is then inserted into the Ward-Takahashi identity. By applying the LSZ reduction procedure to all field insertions, we obtain an equation relating to the scattering amplitude. An algorithmic representation is shown in Fig \ref{fig:route}.

\begin{figure}
    \centering
\[ \lim_{\varepsilon \rightarrow 0}\left(\rule{0em}{12mm}\lim_{\vec{p}\rightarrow 0}
\begin{tikzpicture}[baseline=(m.base)]
    \begin{feynman}
      \vertex[draw,circle,fill=blue!20,minimum size=1.3cm] (m) at ( 0, 0) {$\mathcal{A}^{n+1}(p,\varepsilon)$};
      \vertex (a) at (-1.5,-1) {};
      \vertex (b) at ( 1,-1.5) {};
      \vertex (c) at (-1.5, 1) {};
      \vertex (d) at ( 1, 1.5) {};
      \vertex (e) at ( -0.3, 1.1) {$\cdots$};
      \vertex (f) at ( -0.3, -1.1) {$\cdots$};
      \vertex (g) at ( 1.5, 0) {$p$};
      \diagram* {
        (a) -- [fermion,edge label=] (m) -- [fermion,edge label=] (c),
        (b) -- [fermion,edge label'=] (m) -- [fermion,edge label'=] (d),
        (m) -- [fermion, red] (g)
      };
    \end{feynman}
  \end{tikzpicture} |_{\varepsilon \;\;\text{fixed}}\right)\rule{0em}{12mm}
\]
    \caption{A diagram representation for the soft hierarchy. The $(n+1)$-point soft scattering amplitude is explicitly dependent on the soft momentum $p$ and the integral regulator $\varepsilon$. Here the soft hierarchy implies we keep $\varepsilon$ fixed while taking the soft limit, then take $\varepsilon$ to zero in the final step.} 
    \label{fig:OrderOfLim}
\end{figure}
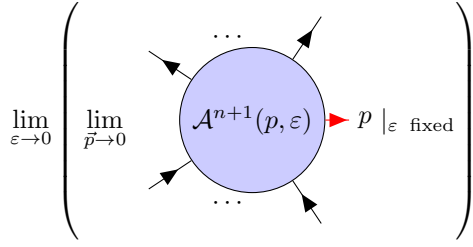

As a pedagogical review, we show in Section \ref{Sec2} how linear Poincar\'{e} symmetry constrains the amplitude following the above route. In the case of soft theorems, there are two implicit limits one needs to impose for the second arrow in the above diagram: soft limit and on-shell limit. The on-shell limit from the LSZ reduction projects out terms regular in on-shell limit, while the soft limit causes higher-order field-dependent terms to drop out of the Noether current. In \cite{Green:2022slj}, it is argued that switching between the order of the two limits is equivalent to transitioning between perturbative and non-perturbative pictures, here perturbative means tree-level. However, the $S$-matrix or the scattering amplitude is a distributional function, thus there is another implicit limit $\varepsilon\rightarrow 0$. The parameter $\varepsilon$ arises from regulating the asymptotic integral. The $\varepsilon$ dependence appears within the energy momentum conserving delta functions $\delta^D(\sum p^\mu)$, LSZ poles $\frac{1}{p^2}$, propagators and other possible residues that cancel the poles. In this work we introduce a new order of limits, namely taking the distributional $\varepsilon\rightarrow 0$ at the very last step after both other two limits, and we refer to this as the \textit{soft hierarchy}. A graphical representation for the soft hierarchy is shown in Fig \ref{fig:OrderOfLim}. The $\varepsilon$ which may be considered as an IR regulator, manifestly commutes the on-shell limit and soft limit, resulting in no dependence on cubic vertices that are not constrained by symmetry, therefore no distinction between perturbative picture and non-pertubative picture argued in \cite{Green:2022slj}. We give a detailed derivation in Section \ref{Sec3} and \ref{Sec5} where we show nothing in the symmetry current contributes to the the soft theorem beyond quadratic order in $\pi$ upon the above \textit{soft hierarchy}. The soft theorem suggests any amplitude that has a soft leg associated with a cubic vertex featuring a co-linear pole from exchanging massless fields does not contribute non-trivially to the soft theorem, exhibiting a \textit{hidden enhanced Adler zero}. Crucially, the $i\varepsilon$ shift is applied exclusively to terms that are indeterminate, or divergent in the limits being considered. Such prescription would ensure the field basis independence of the soft theorems. As an example, in Section \ref{Sec4} we show any diagram derived from the vertex $\dot\pi^3$ satisfies the soft theorem for $\delta\pi = b_i x^i$ derived in \cite{Du:2024hol}
\begin{align}
    \lim_{p\rightarrow 0}\partial_{\vec{p}} \;(\tilde{\mathcal{A}}_{E_p} + \tilde{\mathcal{A}}_{-E_p}) = 0.
\end{align}
 In Section \ref{Sec5} and \ref{Sec6}, we derive soft theorems for non-linear Lorentz boosts and dilatations, demonstrating that they determine the superfluid EFT up to known degrees of freedom in the Wilson coefficients. We then compare the soft theorem constraints to the Hamiltonian Wilson coefficients given in Appendix \ref{LagHam} up to 5-point, and find agreement. Interestingly, in the case of scaling superfluid \cite{Pajer:2018egx,Grall:2020ibl}, where one breaks the linearly realised dilatation symmetry by a time dependent background and the Nambu-Goldstone mode non-linearly realises the symmetry, we're even able to use soft theorem to fix the 3-point amplitude by tweaking the two point amplitude such that the delta structure on both sides of the soft theorem match. In Section \ref{Sec7}, we show the soft theorem contributions from diagrams that are singular in the co-linear limit and contain a cubic soft vertex vanish for both non-linear boost and non-linear dilatation symmetries when summed over all channels, yielding a soft scaling of $\mathcal{O}(p_1^2)$. We then re-emphasise that the soft theorem is field redefinition invariant under the minimal basis if the $i\varepsilon$ shift is only to terms that are divergent or indeterminate in the limit under consideration.

 \paragraph{Summary of results}

The soft theorem for the superfluid EFT, which arises from the flat-space and decoupling limts of the EFT of inflation, with non-linear boost symmetry $\delta_B \pi = b_i[x^i + (x^i\partial_t + t\partial_i)\pi]$ upon canonical normalisation $t\rightarrow t/c_s,\;\pi\rightarrow \sqrt{c_s}\pi$ is
\begin{keyeqn}
\begin{align}
    \lim_{\vec{p}\rightarrow{}0}  \biggl\{ \partial_{p^i} \left [\frac{\tilde{\mathcal{A}}^{n+1}_{E_{p}} + \tilde{\mathcal{A}}^{n+1}_{-E_{p}}}{2}\right]   \biggr\}&= -i \frac{1}{\sqrt{c_s}}\left[\sum_a E_a\partial_{p_a^i}\tilde{\mathcal{A}}^n\right] \,,\\
\lim_{\vec{p}\rightarrow 0} \biggl\{\tilde{\mathcal{A}}^{n+1}_{E_{p}} + \tilde{\mathcal{A}}^{n+1}_{-E_{p}} \biggr\} & =0\,.\label{AdlerZero}
\end{align}
\end{keyeqn}
The second equation is Adler zero which reflects the underlying $U(1)$ shift symmetry of the theory. The soft theorems for scaling superluid, which is a special superfluid that enjoys an additional non-linear dilatation symmetry $\delta_D \pi = -t(\Delta + 1) - (\Delta + x^\mu \partial_{\mu})\pi$, along with non-linear Lorentz boost upon canonical normalisation $t\rightarrow \sqrt{2\alpha-1} t,\;\pi\rightarrow\pi/\sqrt{2\alpha\sqrt{(2\alpha-1)}}$, are 
\begin{keyeqn}
\begin{align}
    \lim_{\vec{p}\rightarrow 0} \biggl\{ \frac{\tilde{\mathcal{A}}^{n+1}_{E_{p}}-\tilde{\mathcal{A}}^{n+1}_{-E_{p}}}{2E_{p}}  \biggr\} & = -i\sqrt{\frac{c_s^5}{1+c_s^2}}\biggl\{- 2(n-2)\biggl(\frac{1-c_s^2}{c_s^2}\biggr) + \frac{1}{\Delta + 1}\bigl[n(-\Delta + D - 2) - D + 
 \vec{p}_a\cdot \partial_{\vec{p}_a}\bigr]\biggr\}  \tilde{\mathcal{A}}^n ,\\
    \lim_{\vec{p}\rightarrow{}0}  \biggl\{ \partial_{p^i} \left [\frac{\tilde{\mathcal{A}}^{n+1}_{E_{p}} + \tilde{\mathcal{A}}^{n+1}_{-E_{p}}}{2}\right]   \biggr\}&= -i \sqrt{\frac{c_s}{1+c_s^2}}\left[\sum_a E_a\partial_{p_a^i}\tilde{\mathcal{A}}^n\right] \,,\\
\lim_{\vec{p}\rightarrow 0} \biggl\{\tilde{\mathcal{A}}^{n+1}_{E_{p}} + \tilde{\mathcal{A}}^{n+1}_{-E_{p}} \biggr\} & =0\,.
\end{align}
\end{keyeqn}
The tilde on the on-shell $(n+1)$-point soft amplitude $\mathcal{A}^{n+1}_{E_p}$ and $n$-point hard amplitude $\mathcal{A}^n$ indicates that the energy momentum conserving delta functions have already been imposed, so no delta function remains within the amplitude. Meanwhile $\mathcal{A}^{n+1}_{-E_p}$ is the same function as $\mathcal{A}^{n+1}_{E_p}$ but with $E_p \rightarrow - E_p$, where $E_p$ is the energy of the particle being taken soft. $c_s$ in both sets of equations are the 'sound speed' parameter, $\Delta$ for scaling superfluid is the scaling dimension of the scalar field, $\alpha$ is related to the space-time dimension $D$ and scaling dimension by
$ \Delta = \frac{D}{2\alpha} - 1$, the 'sound speed' parameter can be written in terms of $\alpha$ by $c_s = \frac{1}{\sqrt{2\alpha - 1}}$. The scattering amplitudes are computed with the $i\varepsilon$ terms explicitly included for terms that are either indeterminate or divergent under soft limit, and they are taken to zero after the soft limit.

We've shown the soft theorems in principal determine all Wilson coefficients given the 'sound speed' parameter $c_s$ \footnote{Since the free theory is canonically normalised, the propagation speed is set to $1$ and the 'sound speed' $c_s$ manifests as a parameter characterising the interaction.}. This has been explicitly checked up to 5-point amplitude.
%%%%%%%%%%%%%%%%%%%%%
\begin{figure}
    \centering
\[ \tilde{\mathcal{A}}_{E_{p}} = 
\begin{tikzpicture}[baseline=(m.base)]
    \begin{feynman}
      \vertex[draw,circle,fill=blue!20,minimum size=1.3cm] (m) at ( 0, 0) {};
      \vertex (a) at (-1.5,-1) {};
      \vertex (b) at ( 1,-1.5) {};
      \vertex (c) at (-1.5, 1) {};
      \vertex (d) at ( 1, 1.5) {};
      \vertex (e) at ( -0.3, 0.8) {$\cdots$};
      \vertex (f) at ( -0.3, -0.8) {$\cdots$};
      \vertex (g) at ( 1.3, 0) {$p$};
      \diagram* {
        (a) -- [fermion,edge label=$p_{r+1} $] (m) -- [fermion,edge label=$p_{r}$] (c),
        (b) -- [fermion,edge label'=$p_{n}$] (m) -- [fermion,edge label'=$p_{1} $] (d),
        (m) -- [fermion, red] (g)
      };
    \end{feynman}
  \end{tikzpicture} 
,\;\;\;\qquad \tilde{\mathcal{A}}_{-E_{p}} = 
  \begin{tikzpicture}[baseline=(m.base)]
    \begin{feynman}
      \vertex[draw,circle,fill=blue!20,minimum size=1.3cm] (m) at ( 0, 0) {};
      \vertex (a) at (-1.5,-1) {};
      \vertex (b) at ( 1,-1.5) {};
      \vertex (c) at (-1.5, 1) {};
      \vertex (d) at ( 1, 1.5) {};
      \vertex (e) at ( -0.3, 0.8) {$\cdots$};
      \vertex (f) at ( -0.3, -0.8) {$\cdots$};
      \vertex (g) at ( 1.4, 0) {$p'$};
      \diagram* {
        (a) -- [fermion,edge label=$p_{r+1} $] (m) -- [fermion,edge label=$p_{r}$] (c),
        (b) -- [fermion,edge label'=$p_{n}$] (m) -- [fermion,edge label'=$p_{1} $] (d),
        (m) -- [fermion, red] (g)
      };
    \end{feynman}
  \end{tikzpicture} 
\]
    \caption{A graphical representation for the energy flipping scattering amplitude. $\tilde{\mathcal{A}}_{E_{p}}$ on the LHS represents an $(n+1)$-point amplitude with $n$ hard momenta denoted by $p_{1}$ to $p_{n}$ and one soft momentum denoted by $p = (E_{p}, \Vec{p})$. The tildes imply that we have imposed energy and momentum conservation by going to minimal basis \cite{Cheung:2023qwn}. $\tilde{\mathcal{A}}_{-E_{p}}$ on the RHS represents the same amplitude but with the energy of the soft momentum flipped i.e. $p'=(-E_{p},\Vec{p})$.} 
    \label{fig:GraphicRep}
\end{figure}
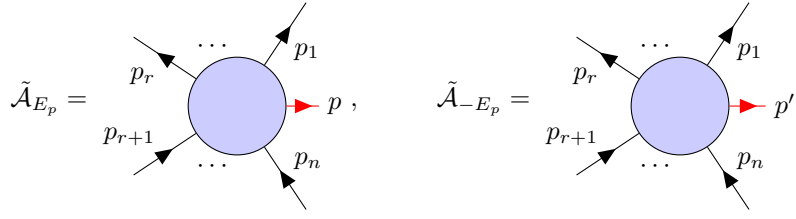

\paragraph{Assumptions and Notations}
Although in the main text we might reiterate some definitions and notations, we list the main assumptions and notations here:
\begin{itemize}
    \item We set the propagation speed to one which is always achievable by rescaling coordinates for linear dispersion relation in the case of a single scalar theory. Therefore, the dispersion relation for massless particles are $E_a \equiv |\vec{p}_a|$, $E_p \equiv |\vec{p}|$.
    \item The Feynman propagator for massless scalar exchange is chosen to be
    \begin{align}
    \Delta(p^0, \vec{p})=\frac{1}{(p^0 - E_p +i\varepsilon)(p^0 + E_p - i\varepsilon)} = \frac{1}{(p^0)^2 - E_p^2 + 2i E_p \varepsilon + \mathcal{O}(\varepsilon^2)},
\end{align}
such that it picks out the correct time ordering. If the propagator is associated with a 3-point vertex where two of which are external, then we have
\begin{align}
    \Delta(E_1 + E_2, \vec{p}_1 + \vec{p}_2) =& \frac{1}{(E_1 + E_2)^2 - (\vec{p}_1 + \vec{p}_2)^2 + 2i\varepsilon E_{p_1 + p_2} +\mathcal{O}(\varepsilon^2)}\nonumber\\
    =& \frac{1}{ 2E_1 E_2 - 2\vec{p}_1\cdot\vec{p}_2 + 2i\varepsilon E_{p_1 + p_2} +\mathcal{O}(\varepsilon^2)}.
\end{align}
The $i\varepsilon$ is only applied when the amplitude is indeterminate or divergent in the limit under consideration.
\item $\bra{0}$ and $\ket{0}$ denote the interaction vacuum states and they are Lorentz invariant even if the background and interaction may break Lorentz invariance, i.e. the Lorentz generators $L_{\mu\nu}$ annihilate the interaction vacuum state
\begin{align}
    L^{\mu\nu}\ket{0} = \bra{0}L^{\mu\nu} = 0.
\end{align}
\item The $S$-matrix for all $n$ particles out-going is denoted as $\braket{\alpha|0} = \braket{\vec{p}_1,\ldots,\vec{p}_n | 0}$, the hard $n$ pt scattering amplitude is parameterised as
\begin{align}
    \braket{\alpha|0} = \braket{\vec{p}_1,\ldots,\vec{p}_n | 0} = \tilde{\mathcal{A}}^{n} \delta\biggl(\sum_a E_a \equiv E\biggr) \delta^d \biggl(\sum_a \vec{p}_a \equiv \vec{p} \biggr),
\end{align}
Here $d$ denotes spatial dimension, and the $\tilde{}$ represents that all the energy momentum conserving delta functions are imposed, in practice it simply means we go to minimal basis (outlined below).  We also use the shorthand notation $E \equiv\sum_a E_a $ and $\vec{p} \equiv \sum_a \vec{p}_a$ to represent the sum of all particles momenta. The soft $(n+1)$-point scattering amplitude is
\begin{align}
    \braket{\alpha,q|0} = \braket{\vec{p}_1,\ldots,\vec{p}_n,\vec{q} | 0} = \tilde{\mathcal{A}}^{n+1}_{E_q} \delta(E + E_q) \delta^d(\vec{p} +\vec{q}),
\end{align}
the energy flipping amplitude is then
\begin{align}
    \braket{\alpha|q'} = \braket{\vec{p}_1,\ldots,\vec{p}_n | -\vec{q}} = \tilde{\mathcal{A}}^{n+1}_{-E_q} \delta(E - E_q) \delta^d(\vec{p} +\vec{q}).
\end{align}
A graphic representation is shown in Fig \ref{fig:GraphicRep}.
\item The conjugate momentum of the field is given by $\tilde{\pi}$ while the time derivative on the field is given by $\dot\pi$.

\item The scattering amplitude is computed such that the on-shell conditions and energy-momentum conservation are imposed. In the case of isotropic background, the scattering amplitude is $SO(D-1)$ invariant where $D$ is the space-time dimension. The on-shell $D$-momentum for an $n$-point amplitude is given by
\begin{align}
    p_a^\mu = (E_a, \vec{p}_a) \;\;\text{for} \;\;1\leq a\leq n.
\end{align}
The invariant building blocks under the constraints are
\begin{keyeqn}
\begin{align}
    E_{a} \;\; &\text{for} \;\;1 \leq a \leq n-1 \,,\\
    s_{b,c} = E_{b} E_{c} - \vec{p}_{b}\cdot \vec{p}_{c} \;\; &\text{for} \;\;1 \leq b < c \leq n-2 \;\;\text{and}\;\; 1 \leq b \leq n-3,\;c = n-1 \,.
\end{align}
\end{keyeqn}
A detailed derivation of this minimal basis can be found in \cite{Du:2024hol,Cheung:2023qwn}.
\end{itemize}

%%%%%%%%%%%%%%%%%%%%%%%%%%%%%%%%%%%%%%%%%%%%%%%%%%%%%
\section{Linearly-realised symmetries and scattering amplitudes}\label{Sec2}
Linearly realised Poincare symmetries impose non-trivial constraints on the $S$-matrix: the $S$-matrix can always be factorised to a distribution product of scattering amplitude and 4-momentum conserving delta function, and the scattering amplitude could only involve Lorentzian product.

In this section we give a heuristic review on how to derive $S$-matrix constraints from linearly realised symmetries by taking explicit care with energy-momentum derivatives and on-shell conditions. Subsequently, we demonstrate how to generalise the same technique to non-linearly realised symmetries which is our main focus and explain why we arrive at soft theorems in that case. 

We start the derivation by writing down the Ward-Takahashi identity for correlation functions arsing from a global symmetry $\pi(x)\rightarrow \pi(x) + \alpha\delta\pi(x)$,
\begin{equation} 
    \partial_\mu\braket{0|\mathcal{T}\{J^\mu(x)\pi(x_1) \ldots \pi(x_n)\}|0} = -  i\sum_{a=1}^{n}\delta^{(D)}(x-x_a)\braket{0|\mathcal{T}\{\pi(x_1) \ldots\delta_a\pi(x_a)\ldots\pi(x_n)\}|0},\label{WardTakahashi}
\end{equation}
where $\partial_\mu$ acts on coordinates $x^{\mu}$, $\mathcal{T}$ represents the usual time-ordering and $\bra{0}$ denotes the interaction vacuum. If we integrate $x$ over the entire space time, the LHS is thus a boundary term. In this section we restrict to the following symmetries that are linearly 
 realised in a Poincare invariant theory:
\begin{align}
    \delta_{\text{translation}}\pi(x) &= a^\mu \partial_\mu \pi(x),\\
    \delta_{\text{Boost}}\pi(x) &= b^i (t\partial_i + x^i\partial_t) \pi(x),\\
    \delta_{\text{Rotation}}\pi(x) &= r^{ij} (x_i\partial_j - x_j\partial_i) \pi(x),
\end{align}
where $a^\mu, b^{i}, r^{ij}$ are global parameters. In this case, Poincare symmetries are clearly preserved by the asymptotic boundary, thus the boundary term vanishes on the LHS when acted upon the vacuum
\begin{align}
    \bra{0}\int_{\partial D}dS_\mu J^\mu = \int_{\partial D}dS_\mu J^\mu \ket{0} = 0.
\end{align}
After the integration, equation \eqref{WardTakahashi} becomes
\begin{align} \label{WTLinear}
    0 = \sum_{a=1}^{n} \braket{0|\mathcal{T}\{\pi(x_1) \ldots\delta\pi(x_a)\ldots\pi(x_n)\}|0}.
\end{align}
To arrive at constraints for on-shell scattering amplitude, we perform LSZ reduction, and for simplicity we put all particles in the out state such that \eqref{WTLinear} becomes
\begin{align}
    0 =   \prod_{a=1}^n \lim_{p_a^0 \rightarrow E_a} {p_a^2} \int_{x_a} e^{i p_a\cdot x_a} \sum_{b = 1}^n   \braket{0|\mathcal{T}\{\pi(x_1) \ldots\delta_b\pi(x_b)\ldots\pi(x_n)\}|0}. \label{LinearLSZ}
\end{align}
where we've defined $p_a\cdot x_a \equiv p_a^0 t_a -\Vec{p}_a\cdot\vec{x}_a$ and ${p_a^2}\equiv (p_a^0 + E_a)(p_a^0 - E_a )$. Now we discuss the three cases above separately. 

\subsection{Space-time translations}
To perform LSZ reduction, we split every time integral in \eqref{LinearLSZ} into three regions
\begin{align}
    \int_{x_a} e^{i p_a \cdot x_a} \ldots= \biggl[\int_{t_a>T^+} + \int_{T^-\leq t_a\leq T^+} + \int_{t_a<T^-} \biggr] \int_{\vec{x}_a} e^{ip_a\cdot x_a} \ldots,
\end{align}
here the $\ldots$ denotes contributions from the correlation functions. Then \eqref{LinearLSZ} can be re-written as
\begin{align}
    0 = \prod_{a}^n\biggl[G_a^+ + G_a^0 + G_a^-\biggr],
\end{align}
where
\begin{align}
    G_a^+ &= \lim_{p_a^0 \rightarrow E_a} {p_a^2}\int_{t_a> T^+}\int_{\vec{x}_a} e^{i p_a\cdot x_a} \sum_b   \braket{0|\mathcal{T}\{\pi(x_1) \ldots\delta_b\pi(x_b)\ldots\pi(x_n)\}|0},\\
    G_a^0 &= \lim_{p_a^0 \rightarrow E_a} {p_a^2} \int_{T^-\leq t_a\leq T^+}\int_{\vec{x}_a} e^{i p_a\cdot x_a} \sum_b   \braket{0|\mathcal{T}\{\pi(x_1) \ldots\delta_b\pi(x_b)\ldots\pi(x_n)\}|0},\\
    G_a^- &= \lim_{p_a^0 \rightarrow E_a} {p_a^2} \int_{t_a< T^-}\int_{\vec{x}_a} e^{i p_a\cdot x_a} \sum_b   \braket{0|\mathcal{T}\{\pi(x_1) \ldots\delta_b\pi(x_b)\ldots\pi(x_n)\}|0}.
\end{align}
First we notice that the intermediate and far past integral $G^0, G^-$  vanish since they do not generate any pole at $p_a^0 = E_a$ due to the specific choice of our limit $p_a^0 \rightarrow E_a$ that is uniquely picking out the residue at $p_a^0 =  E_a$. Since the pole at $p_a^0 = E_a$ is only generated at the asymptotic future, it precisely captures the \textit{out} state contribution, resulting in the all-out scattering amplitude \cite{Peskin:1995ev}. Therefore we only focus on the far future integral $t_a>T^+$, 
\begin{align}
    0 =\prod_a^n G_a^+ =   \prod_{a}^n \lim_{p_a^0 \rightarrow E_a} {p_a^2}\int_{t_a>T^+}\int_{\vec{x}_a} e^{i p_a\cdot x_a} \sum_b \braket{0|\mathcal{T}\{\pi(x_1) \ldots \partial^b_\mu\pi(x_b)\ldots\pi(x_n)\}|0},
\end{align}
here $\partial^b_\mu$ denotes derivative on $x_b$. The integral reads
\begin{align}
    0 =& \prod_{a}^n \lim_{p_a^0 \rightarrow E_a} \sum_b {p}_{b}^\mu|_{\text{on-shell}}\frac{(p_a^0 + E_a)\Ccancel{(p_a^0 - E_a )}e^{i(p_a^0 - E_a +i\varepsilon)T^+}}{2E_a\Ccancel{(p_a^0 - E_a )}} \braket{\alpha|0}\\
    =& e^{-n\varepsilon T^+}\sum_b {p}_{b}^\mu|_{\text{on-shell}} \braket{\alpha|0}. \label{EnergyMomentum}
\end{align}
where we denote ${p}_{b}^\mu|_{\text{on-shell}} = (E_b, \Vec{p}_b) $ as on-shell momenta to distinguish from LSZ parameter $p_a^\mu = (p_a^0, \Vec{p}_a)$. $\braket{\alpha|0}$ is a shorthand for the $S$-matrix element $\braket{\vec{p}_1,\vec{p}_2,\ldots,\vec{p}_n|0}$. The $i\varepsilon$ shift is only applied in the occurrence of indetermination or divergence, terms that are regular in the limit under consideration do not receive the $i\varepsilon$ shift, therefore we're safe to discard the $i\varepsilon$ from the future time integral. The prescription ensures field redefinition invariance of the scattering amplitude equations. We also discard multi-particle state contributions as they do not generate poles at $p_a^0 = E_a$. Since the on-shell amplitude is non-zero, the above relation \eqref{EnergyMomentum} implies that we must have energy and momentum conservation, i.e. $\sum_b {p}_{b}^\mu|_{\text{on-shell}} = 0$. 
From the above equation, it's straightforward to extrapolate,
\begin{align}
    \braket{\alpha|0} \propto \delta(E) \delta^d(\Vec{p}).
\end{align}
Henceforth we parameterise the $S$-matrix by $\braket{\alpha|0} = \mathcal{A}^n \delta(E) \delta^{d}(\Vec{p})$ where $\mathcal{A}^n$ is the scattering amplitude.

\subsection{Spatial rotations}
We now turn our attention to spatial rotations and derive the conditions that they impose on scattering amplitudes. As was the case for translations,  the only contribution comes from far future integral,
\begin{align}
    0 = \prod_{a=1}^n G_a^+ = \prod_{a}^n \lim_{p_a^0 \rightarrow E_a} {p_a^2}\int_{t_a>T^+}\int_{\vec{x}_a} e^{i p_a\cdot x_a} \sum_b   \braket{0|\mathcal{T}\{\pi(x_1) \ldots (x_i\partial_j - x_j\partial_i)_b\pi(x_b)\ldots\pi(x_n)\}|0}.
\end{align}
Notice that the Fourier dual of $x^b_i$, $p_b^i$, is off-shell. We have
\begin{align}
    0 =    e^{-(n-1)\varepsilon T^+}   \sum_a \lim_{p_a^0\rightarrow E_a}{p_a^2} (p_{a}^j\partial_{p_a}^i- p_{a}^i\partial_{p_a}^j)\biggl[\frac{e^{i(p_a^0 - E_a +i\varepsilon)T^+} \braket{\alpha|0}}{2E_a(p_a^0 - E_a + i\varepsilon)}\biggr].
\end{align}
 We explicitly include the $i\varepsilon$ since each of the two terms is divergent in $p_a^0 \rightarrow E_a$, however they cancel against each other. Expanding out the derivatives, not surprisingly, every other term cancels besides
\begin{align}
     0 = \sum_a (p_a^j \partial_{p_a}^i - p_a^i \partial_{p_a}^j) \braket{\alpha|0}.
\end{align}
The equation implies the $S$-matrix is rotation invariant. To show that the scattering amplitude is also rotation invariant, we insert the delta-function factorisation from the last subsection i.e. $\braket{\alpha|0} = \mathcal{A}^n \delta(E) \delta^{d}(\Vec{p})$ into the above equation yielding
\begin{align}
    0 =& \sum_a (p_a^j \partial_{p_a}^i - p_a^i \partial_{p_a}^j) \left[\mathcal{A}^n \delta(E) \delta^d(\Vec{p})\right]\nonumber\\
    =& \sum_a [(p_a^j \partial_{p_a}^i - p_a^i \partial_{p_a}^j)\mathcal{A}^n]\delta(E) \delta^d(\vec{p}) + [(p_a^j \partial_{p_a}^i - p_a^i \partial_{p_a}^j)\delta(E)] \mathcal{A}^n\delta^d(\vec{p}) \nonumber\\
    &+ [(p_a^j \partial_{p_a}^i - p_a^i \partial_{p_a}^j)\delta^d(\vec{p})] \mathcal{A}^n\delta(E) \nonumber\\
    =& \sum_a [(p_a^j \partial_{p_a}^i - p_a^i \partial_{p_a}^j)\mathcal{A}^n]\delta(E) \delta^d(\vec{p}) + \biggl[\frac{(p_a^j p_a^i - p_a^i p_a^j)}{E_a}\delta'(E)\biggr] \mathcal{A}^n\delta^d(\Vec{p})\nonumber\\
    &+ [\partial_{p_a}^i (p_a^j\delta^d(\vec{p})) - \partial_{p_a}^j (p_a^i\delta^d(\vec{p}))]\mathcal{A}^n\delta(E) - (\eta^{ji} - \eta^{ij})\mathcal{A}^n \delta(E) \delta^d(\vec{p}) \nonumber\\
    =& \sum_a [(p_a^j \partial_{p_a}^i - p_a^i \partial_{p_a}^j)\mathcal{A}^n]\delta(E) \delta^d(\vec{p}),
\end{align}
where $\eta^{\mu\nu} = \text{diag}(1,-1,\ldots,-1)$. We imposed momentum conservation $\sum_a p_a^j \delta^d(\Vec{p}) = 0$ to arrive at the last line. The reason we could do the summation first before taking derivatives is because $\partial_{p_a}^i\delta^d(\Vec{p})$ is the same for every $a$ since they're symmetric within the delta function. Therefore we arrive at the scattering amplitude constraint from rotation symmetry
\begin{align}
    \sum_a (p_a^j \partial_{p_a}^i - p_a^i \partial_{p_a}^j)\mathcal{A}^n = 0. \label{rotation}
\end{align}
The rotation constraint \eqref{rotation} is sufficient to conclude that only three-momentum contractions are allowed in the scattering amplitude.

\subsection{Lorentz boosts}
Finally, for the linear realisation of Lorentz boosts, we have
\begin{align}
    0&= \prod_{a=1}^n G_a^+ =   \prod_{a=1}^n \lim_{p_a^0 \rightarrow E_a} {p_a^2} \int_{t_a>T^+}\int_{\vec{x}_a} e^{i p_a\cdot x_a} \sum_{b=1}^n  \braket{0|\mathcal{T}\{\pi(x_1) \ldots (t\partial_i + x^i\partial_t)_b\pi(x_b)\ldots\pi(x_n)\}|0}\nonumber\\
    &= e^{-(n-1)\varepsilon T^+}\sum_a \lim_{p_a^0\rightarrow E_a}{p_a^2}\biggl\{\partial_{p_a^0} \biggl[\frac{p_{a}^i  e^{i(p_a^0 - E_a +i\varepsilon)T^+} \braket{\alpha|0}}{2E_a(p_a^0 - E_a + i\varepsilon)}\biggr] +  \partial_{p_a^i} \biggl[\frac{ E_a e^{i(p_a^0 - E_a +i\varepsilon)T^+} \braket{\alpha|0}}{2E_a(p_a^0 - E_a + i\varepsilon)}\biggr]\biggr\} \nonumber\\
    &= e^{-n\varepsilon T^+} \sum_{a}\biggl[i\biggl(T^+ + \frac{1}{\varepsilon}\biggr)(p_a^i - p_a^i) + E_a\partial_{p_a^i}\biggr]\braket{\alpha|0}. \label{LIBoost}
\end{align}
Again the $i\varepsilon$ is explicitly included but the divergence cancels each other so including it doesn't make a difference. The $S$-matrix equation is then
\begin{align}
    0 = \sum_a E_{a}\partial_{p_a^i}\braket{\alpha|0},
\end{align}
if we keep $i\varepsilon$ fixed and send $p_a^0 \rightarrow E_a$. We therefore have
\begin{align}
    0&=\sum_a E_{a}\partial_{p_a^i}\braket{\alpha|0}\nonumber\\ 
    &= \sum_a E_a \partial_{p_a^i}\left[{\mathcal{A}}^n \delta(E)\delta^d(\Vec{p})\right]\nonumber\\
    &= \sum_a [E_a \partial_{p_a^i}{\mathcal{A}}^n] \delta(E)\delta^d(\vec{p}) - \sum_a [p_a^i\delta'(E)]{\mathcal{A}}^n\delta^d(\Vec{p}) + \sum_a [E_a \partial_{p_a^i}\delta^d(\vec{p})] \tilde{\mathcal{A}}^n \delta(E)\nonumber\\
    &= \sum_a [E_a \partial_{p_a^i}{\mathcal{A}}^n] \delta(E)\delta^d(\Vec{p}),
\end{align}
where to arrive at the last line we impose energy and momentum conservation to eliminate the last two terms from the second last line. We notice that $\partial_{p_a^i}\delta^d(\vec{p})$ is the same for every $a$ since they're symmetric within the delta function\footnote{The argument is not admissible for two point amplitudes where the energy and momentum delta function are not independent. As we shall see in Section \ref{Sec6}, we come up with a Lorentz invariant two point function that matches the soft theorem delta structure.}. Therefore we arrive at the scattering amplitude equation 
\begin{align}
    0 = \sum_a E_a \partial_{p_a^i}{\mathcal{A}}^n. \label{boost}
\end{align}
The boost constraint \eqref{boost} together with rotation constraint \eqref{rotation}
are sufficient to conclude only four-momentum contractions are allowed in scattering amplitude. For instance, let's consider the Lorentzian product $p_1\cdot p_2 =E_1 E_2 - \vec{p}_1\cdot \vec{p}_2 \subset \mathcal{A}^n$,
\begin{align}
    (E_1 \partial_{p_1^i} + E_2 \partial_{p_2^i} +\ldots)(E_1 E_2 - \Vec{p}_1\cdot \Vec{p}_2) = p_1^i E_2 - p_2^i E_1 + p_2^i E_1 - p_1^i E_2 = 0.
\end{align}
We see by specifying on-shell condition, energy derivative does not appear in the resulting scattering amplitude equation, since our one-particle states completeness relation is parameterised as
\begin{align}
    1 = \int_{\vec
    {p}} \frac{d^3\vec{p}}{2E_p} \ket{\vec{p}}\bra{\vec{p}},
\end{align}
where the energy dependence is eliminated by momentum using on-shell condition $E_p = |\vec{p}|$ for asymptotically free states.

\section{Soft theorem derivation}\label{Sec3}
In this section, we will first explain why the soft limit is special in symmetry-breaking scenarios, followed by a detailed derivation of soft theorems for theories with generic non-linear symmetries. During the derivation, we will highlight the importance of the \textit{soft hierarchy}, i.e. soft momenta $q$ is hierarchically smaller than the integral regulator $\varepsilon$, and demonstrate that the soft theorem is inherently independent of any unconstrained off-shell cubic vertex. 
\subsection{Why are soft limits special?}
\begin{figure}[hbt!]
    \centering
    \includegraphics[width=16cm]{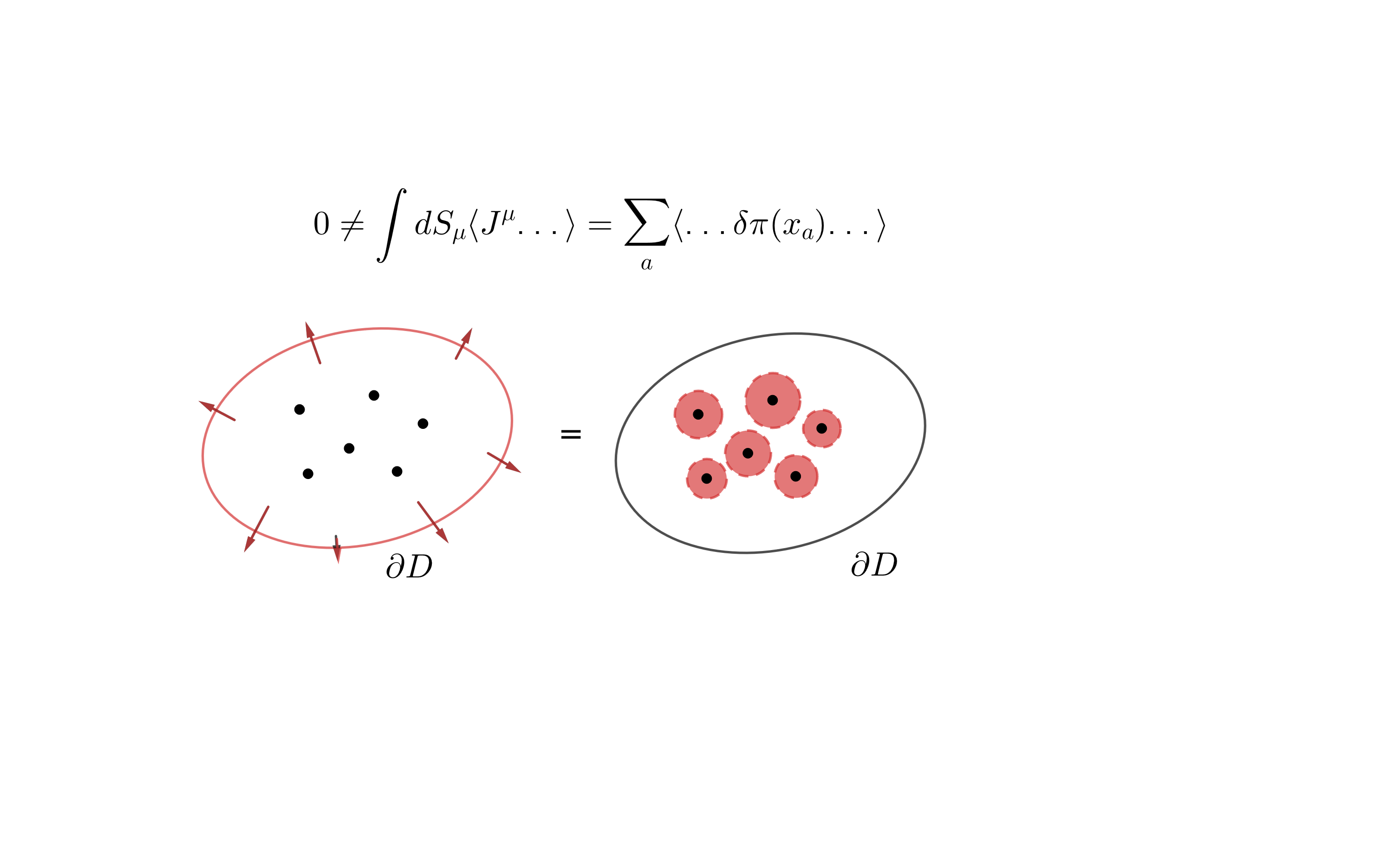}
    \caption{A graphic representation of non vanishing boundary integral of Ward Takahashi identity. When the boundary $\partial D$ is not invariant under the symmetry associated with the current $J^\mu$, the RHS does not vanish.}
    \label{boundary}
\end{figure}

Why are soft limits interesting? One naive answer to the question is that the vacuum symmetry breaking structure is inextricably connected to the IR behavior of scattering amplitudes, therefore the vacuum configuration is encoded in the amplitude of which the momenta is close to zero. To make the statement concrete, let's assume we break the Lorentz boost symmetry by a non-invariant ground state in the context of inflation, $\braket{\phi} = t$, so that boosts are non-linearly realised by the perturbation as \cite{Grall:2020ibl, Green:2022slj, Cheung:2023qwn}
\begin{align}
    \delta_B \pi = b_i[x^i + (x^i\partial_t + t\partial_{i})\pi].
\end{align}
Now we wish to plug this symmetry transformation into the Ward-Takahashi identity \eqref{WardTakahashi} to derive constraints on the corresponding scattering amplitudes of $\pi$ as we did in the last section for the linearly realised symmetries. We quickly run into a problem. Due to the presence of the non-linear term, the current associated with this boost-breaking symmetry does not vanish at the boundary, thus the LHS of Ward-Takahashi identity \eqref{WardTakahashi} after integration is non-zero. The non-zero RHS it given by
\begin{align}
\sum_a\int_{D_a}\braket{0|\mathcal{T}\{\ldots\delta\pi(x_a)\ldots\}|0} = \int_{\partial D} dS^\mu \braket{0|\mathcal{T}\{J^\mu(x)\ldots\}|0}\neq 0.
\end{align}
A graphical representation is shown in Fig \ref{boundary}. The resolution turns out to be Fourier transforming to momentum space $q$. First, we multiply both sides by $e^{i q\cdot x}$. Then, using integration by parts, the space-time derivative $\partial_\mu$ brings down a front factor of $q^\mu$. The boundary term vanishes due to the $i\varepsilon$ prescription. The Noether current $J^\mu$ depends on the interaction, since we're interested in model independent properties of the theory, we send $q\rightarrow 0$ to cancel out the explicit unconstrained off-shell interaction dependence which we would be explicitly calculating in this section,
\begin{align}
    \lim_{q\rightarrow 0} -iq_\mu \int_D e^{iq\cdot x}\braket{0|\mathcal{T}\{J^\mu(x)\ldots\}|0} = -i\sum_{a} e^{iq\cdot x_a}\braket{0|\ldots\delta_a\pi(x_a)\ldots|0}.
\end{align}
To arrive at a constraint on the scattering amplitude, we perform LSZ reduction to pick out the asymptotic contribution. 
\begin{keyeqn}
    \begin{align}
    \lim_{q\rightarrow 0} -iq_\mu \int_D e^{iq\cdot x}\prod_{a=1}^n\text{LSZ}_{a+}\braket{0|\mathcal{T}\{J^\mu(x)\ldots\}|0} = -i\sum_{b} e^{iq\cdot x_b}\prod_{a=1}^n\text{LSZ}_{a+}\braket{0|\ldots\delta_b\pi(x_b)\ldots|0},\label{qWT}
\end{align}
\end{keyeqn}
where the LSZ operator is defined by
\begin{align}
    \text{LSZ}_{a+} = -i\lim_{p_a^0\rightarrow E_a} {p_a^2}
    \int_{x_a}e^{ip_a\cdot x_a}.
\end{align}
We see that the above soft theorem arises due to the non-trivial behavior of the boundary under non-linear transformations and by maintaining control over higher-order terms in momenta. In this section we give a detailed derivation for general non-linear symmetries and show that it is valid for all-order perturbations even in the presence of 3-point vertices. We evaluate the LHS and RHS of \eqref{qWT} individually.
\subsection{LHS of \eqref{qWT}}
We evaluate the LHS of \eqref{qWT}, which is given by
\begin{align}
   \text{LHS} = \lim_{q\rightarrow 0}-i q_\mu \int_{x} e^{iq\cdot x} \prod_{a=1}^n \text{LSZ}_{a+} \braket{0|\mathcal{T}\{J^{\mu}(x)\pi(x_1) \ldots \pi(x_n)\}|0},
\end{align}
by explicitly taking care of the time ordering and $i\varepsilon$ prescription. Let us consider the conserved current for a general Lagrangian along with its global symmetry
\begin{align}
    \partial_\mu J^\mu = \sum_{k=0}^\infty (-1)^k \partial_{\mu_1}\partial_{\mu_2}\ldots\partial_{\mu_k} \frac{\partial \mathcal{L}}{\partial \partial_{\mu_1}\partial_{\mu_2}\ldots\partial_{\mu_k}\pi}\delta\pi.
\end{align}
The symmetry transformation can be split, as done in \cite{Bittermann:2022nfh}, into:
\begin{align}
    \delta\pi = \delta_{(0)}\pi + \delta_{(1)}\pi +\ldots.
\end{align}
This distinguishes among non-linear(sub-linear) $\delta_{(0)}\pi$, linear $\delta_{(1)}\pi$ and higher order in $\pi$ parts of the transformation. We also split the Noether current in the same manner by
\begin{align}
    J^\mu = J_{(1)}^\mu + J_{(2)}^\mu + \ldots,
\end{align}
where $J_{(1)}^\mu$ is linear in $\pi$, $J_{(2)}^\mu$ is quadratic in $\pi$ and $\ldots$ denotes terms that are higher orders in $\pi$. The current starts at linear order as there is no taddpole in the Lagrangian. We discuss in separate cases based on the order of $\pi$ in which the current consists of.
\paragraph{$J^\mu_{(1)}$ contribution: } 
Let us first consider a general spatial polynomial shift symmetry
\begin{align}
    \delta_{(0)}\pi =b_{i_1 i_2 \ldots i_N} x^{i_1} x^{i_2}\ldots x^{i_N}.\label{spatialsym}
\end{align}
We use $N$ to denote the polynomial degree of the spatial shift symmetry and $n$ to denote the number of external particles in a given scattering amplitude. Since EOM starts at linear order in $\pi$, the current at linear order has to be $\text{EOM}\times \delta_{(0)}\pi$ \footnote{The expansion is important such that the derivative $\partial_\mu$is outside the time-ordered correlation.}
\begin{align}
    \partial_\mu J^\mu_{(1)} &= b_{i_1 i_2 \ldots i_N} \Box\pi x^{i_1} x^{i_2}\ldots x^{i_N} \nonumber\\ 
    &= b_{i_1 i_2 \ldots i_N}\biggl[\partial_\mu(\partial^\mu\pi x^{i_1} x^{i_2}\ldots x^{i_N} ) - N \partial^{i_1}(\pi x^{i_2}\ldots x^{i_N}) + N(N-1)\pi \eta^{i_1 i_2} x^{i_3}\ldots x^{i_N} \biggr].
\end{align}
 Here we define the LHS contribution from the current that is linear in $\pi$ by $G_{(1)}$ and split it into three regions
\begin{align}
    G_{(1)} = G_{(1)}^+ + G_{(1)}^0 + G_{(1)}^-,
\end{align}
where 
\begin{align}
    G_{(1)}^+ &= \lim_{q\rightarrow 0} -i q_\mu \int_{t>T^+}\int_{\vec{x}} e^{iq\cdot x} \prod_{a=1}^n \text{LSZ}_{a+} \braket{0|\mathcal{T}\{J^{\mu}_{(1)}(x)\pi(x_1) \ldots \pi(x_n)\}|0}, \\
    G_{(1)}^0 &= \lim_{q\rightarrow 0} -i q_\mu \int_{T^- \leq t\leq T^+}\int_{\vec{x}} e^{iq\cdot x} \prod_{a=1}^n \text{LSZ}_{a+} \braket{0|\mathcal{T}\{J^{\mu}_{(1)}(x)\pi(x_1) \ldots \pi(x_n)\}|0}, \\
    G_{(1)}^- &= \lim_{q\rightarrow 0} -i q_\mu \int_{t<T^-}\int_{\vec{x}} e^{iq\cdot x} \prod_{a=1}^n \text{LSZ}_{a+} \braket{0|\mathcal{T}\{J^{\mu}_{(1)}(x)\pi(x_1) \ldots \pi(x_n)\}|0} .
\end{align}
Inserting $J^\mu_{(1)}$ into the WT identity and performing the LSZ reduction while focusing on the far future integral $G_{(1)}^+$ yields
\begin{align}
   G_{(1)}^+ =& \lim_{q\rightarrow 0}(-i)^{N+3} \biggl\{\biggl(q_\mu \partial_{q_{i_1}}\ldots \partial_{q_{i_N}} \biggr)\biggl[\frac{q|_{\text{on-shell}}^\mu e^{i(q^0 - E_q +i\varepsilon)T^+}\braket{\alpha,q|0}}{2E_{q}(q^0 - E_q +i\varepsilon)}\biggr]\nonumber\\
    +& \biggr[N q^{i_1}\partial_{q_{i_2}}\ldots \partial_{q_{i_N}}  
    + N(N-1) \eta^{i_1 i_2}\partial_{q_{i_3}}\ldots \partial_{q_{i_N}}     \biggr] \biggl[\frac{ e^{i(q^0 - E_q +i\varepsilon)T^+}\braket{\alpha,q|0}}{2E_{q}(q^0 - E_q +i\varepsilon)}\biggr]\biggr\},
\end{align}
where again we denote $q|_{\text{on-shell}}^\mu \equiv (E_q, \Vec{q})$ as on-shell momenta to distinguish from Fourier parameter $q^\mu = (q^0, \Vec{q})$ and the $S$-matrix is parameterised as $\braket{\alpha,q|0} = \mathcal{A}^{n+1}(\vec{p}_1, \Vec{p}_2,\ldots,\vec{p}_n,\vec{q})\delta(E + E_q)\delta^d(\vec{p} + \vec{q})$, we also drop symmetry parameter to avoid clutter. The latter two terms are explicitly canceled by the first term, leaving us with
\begin{align}
G^+_{(1)} &= \lim_{q\rightarrow 0}(-i)^{N+3} \partial_{q_{i_1}}\ldots \partial_{q_{i_N}} \biggl[\frac{q_\mu q|_{\text{on-shell}}^\mu e^{i(q^0 - E_q +i\varepsilon)T^+}\braket{\alpha,q|0}}{2E_{q}(q^0 - E_q )}\biggr]  \nonumber\\
&= \lim_{q\rightarrow 0}  \frac{(-i)^{N+3}}{2}\partial_{q_{i_1}}\ldots \partial_{q_{i_N}} \biggl[e^{i(q^0 - E_q )T^+}\braket{\alpha,q|0}\biggr].\label{spatialderivation}
\end{align}
We drop the $i\varepsilon$ on the pole since there is no divergence nor indetermination. We drop multi-particle state contributions since they remain regular in the soft limit, even when acted upon by momentum derivatives, as the effective mass prevents IR divergence, thus annihilated by the soft momenta in the numerator \cite{Peskin:1995ev}.
For the far past integral, the calculation is similar
\begin{align}
    G^-_{(1)} =  \frac{(-i)^{N+3}}{2}\partial_{q_{i_1}}\ldots \partial_{q_{i_N}} \biggl[e^{i(q^0 + E_q -i\varepsilon)T^-}\braket{\alpha|q'}\biggr],
\end{align}
where $q' = (E_q, -\vec{q})$. Only the soft mode enters the intermediate region since other contributions are projected out by $p_a^0 \rightarrow E_{a} $, we can thus take the continuous limit $T^+\rightarrow T^- = T$ without affecting the result, thus we could discard contribution from $G^0_{(1)}$. Hence we have
\begin{align}
    G_{(1)} &= \lim_{T^-\rightarrow T^+} [G^+_{(1)} + G^-_{(1)}]\nonumber\\
    &= \lim_{q\rightarrow 0} \frac{(-i)^{N+3}}{2}\partial_{q_{i_1}}\ldots \partial_{q_{i_N}} \biggl[e^{i(q^0 - E_q +i\varepsilon)T}\braket{\alpha,q|0} +e^{i(q^0+E_q -i\varepsilon)T}\braket{\alpha|q'}\biggr].
\end{align}
 To proceed, we use the momentum delta function $\delta^{d}(\vec{p} + \vec{q})$ from the $S$-matrix to replace $\vec{q}$ with $\vec{p}$ such that we can commute all the $q$ derivatives with the scattering amplitude $\mathcal{A}^{n+1}(\vec{p}_1, \Vec{p}_2,\ldots,\vec{p}_n,-\vec{p})$. The upshot is that derivatives acting on spatial $\delta^{d}(\vec{p} + \vec{q})$ would generate a tower structure yielding \cite{Du:2024hol}
\begin{align}
    b_{i_{1} \ldots i_{n}}     \lim_{\vec{p}\rightarrow 0} \biggl\{\partial_{p_{i_1}}\dots\partial_{p_{i_k}}\left[\tilde{\mathcal{A}}_{E_{p}}+\tilde{\mathcal{A}}_{-E_{p}} \right] \biggr\} & =\ldots\,,k=0,1,2,\ldots,N\,.\label{spatial}
\end{align}
where $E_p \equiv |\vec{p}| $ and $\ldots$ represents other contributions to the soft theorem from higher order in $\pi$ terms in current and RHS which we will discuss below. Due to the presence of the momentum derivatives on the LHS, we get a series of derivatives of energy-momentum conserving delta functions. Since the $n$th and $m$th derivatives of delta functions, $\delta^{(m)}(x)$ and $\delta^{(n)}(x)$, are linearly independent when $m\neq n$, this gives us a series of equations for the scattering amplitude by matching terms with identical delta function structures. We refer readers to \cite{Du:2024hol} for a more comprehensive discussion of this tower structure. 

When the non-linear symmetry depends on time, namely
\begin{align}
    \delta_{(0)}\pi =c_{i_1 i_2 \ldots i_N} t x^{i_1} x^{i_2}\ldots x^{i_N},
\end{align}
we do not include quadratic and higher order in $t$ terms since they do not generate new constraints \cite{Du:2024hol}. The current at linear order in $\pi$ is
\begin{align}
    &\partial_\mu J^\mu_{(1)} \nonumber\\
     =& c_{i_1 i_2 \ldots i_N} \Box\pi t x^{i_1} x^{i_2}\ldots x^{i_N}\nonumber\\
    =& c_{i_1 i_2 \ldots i_N} \biggl[\partial_\mu (\partial^\mu\pi t x^{i_1} x^{i_2}\ldots x^{i_N}) - \partial_t(\pi x^{i_1} x^{i_2}\ldots x^{i_N}) - N\partial^{i_1}(\pi t x^{i_2}\ldots x^{i_N}) + N(N-1)\eta^{i_1 i_2} t x^{i_3}\ldots x^{i_N} \biggr].
\end{align}
In this case the far future integral $G^+_{(1)}$ takes the form of
\begin{align}
    G^+_{(1)} &= \lim_{q\rightarrow 0}(-i)^{N} \biggl\{\biggl(q_\mu \partial_{q_{i_1}}\ldots \partial_{q_{i_N}}\partial_{q_0}\biggr) \biggl[\frac{q|_{\text{on-shell}}^\mu e^{i(q^0 - E_q +i\varepsilon)T}\braket{\alpha,q|0}}{2E_{q}(q^0 - E_q +i\varepsilon)}\biggr] \nonumber\\
    &+ \biggl[q^{0}\partial_{q_{i_1}}\ldots \partial_{q_{i_N}}  
    + n q^{i_1}\partial_{q_{i_2}}\ldots \partial_{q_{i_N}}\partial_{q_0}
    + n(n-1) \eta^{i_1 i_2}\partial_{q_{i_3}}\ldots \partial_{q_{i_N}}\partial_{q_0}  \biggr] \biggl[\frac{e^{i(q^0 - E_q +i\varepsilon)T}\braket{\alpha,q|0}}{2E_{q}(q^0 - E_q +i\varepsilon)}\biggr]\biggr\}.
\end{align}
The cancellation among these terms becomes subtle.
After simplification, the future integral is then
\begin{align}
    G^+_{(1)} &= (-i)^N\lim_{q\rightarrow 0} \biggl\{\partial_{q_{i_1}}\ldots \partial_{q_{i_N}}\biggl[\frac{(q^0-q|_{\text{on-shell}}^0)e^{i(q^0 - E_q +i\varepsilon)T}\braket{\alpha,q|0}}{2E_q (q^0 - E_q)}\biggr] \nonumber\\
    &+ \partial_{q_{i_1}}\ldots \partial_{q_{i_N}} \biggl[\frac{q_\mu q|_{\text{on-shell}}^\mu e^{i(q^0 - E_q +i\varepsilon)T^+}\braket{\alpha,q|0}}{2E_{q}(q^0 - E_q)}\biggr]\biggr\}\nonumber\\
    &= (-i)^N \lim_{q\rightarrow 0} \biggl\{\partial_{q_{i_1}}\ldots \partial_{q_{i_N}} \biggl[e^{i(q^0 - E_q +i\varepsilon)T}\frac{\braket{\alpha,q|0}}{2E_q}\biggr] + \partial_{q_{i_1}}\ldots \partial_{q_{i_N}} \biggl[e^{i(q^0 - E_q +i\varepsilon)T^+}\braket{\alpha,q|0}\biggr]\biggr\}.
\end{align}
Again, the $i\varepsilon$ is dropped due to the absence of divergence or indetermination in the soft limit. Together with its corresponding term in the far past yields the soft theorem structure found in \cite{Du:2024hol} is
\begin{align}
    G_{(1)} = &\lim_{q\rightarrow 0} (-i)^N\partial_{q_{i_1}}\ldots \partial_{q_{i_N}} \biggl[ e^{i(q^0 - E_q +i\varepsilon)T}\frac{\braket{\alpha,q|0}}{2E_q} - e^{i(q^0+E_q -i\varepsilon)T}\frac{\braket{\alpha|q',0}}{2E_q} \biggr] \nonumber\\
    +&\lim_{q\rightarrow 0} \frac{(-i)^{N}}{2}\partial_{q_{i_1}}\ldots \partial_{q_{i_N}}\partial_{q_0} \biggl[e^{i(q^0 - E_q +i\varepsilon)T}\braket{\alpha,q|0} +e^{i(q^0+E_q -i\varepsilon)T}\braket{\alpha|q',0}\biggr].
\end{align}
The soft theorem contribution is then
\begin{align}
    c_{i_{1} \ldots i_{N}} \lim_{\vec{p}\rightarrow 0} \biggl\{\partial_{p_{i_1}}\dots\partial_{p_{i_k}}\left[ \tilde{\mathcal{A}}_{E_{p}}+ \tilde{\mathcal{A}}_{-E_{p}} \right] \biggr\} & = \ldots \,,\;\;\;k=0,1,2,\ldots,N \,,  \\
c_{i_{1} \ldots i_{N}} \lim_{\vec{p}\rightarrow 0} \biggl\{\partial_{p_{i_1}}\dots\partial_{p_{i_k}}\left[\frac{\tilde{\mathcal{A}}_{E_{p}}-\tilde{\mathcal{A}}_{-E_{p}}}{E_{p}} \right] \biggr\} & =\ldots \,,\;\;\;k=0,1,2,\ldots,N \,.
\end{align}
Again the equal sign is obtained by comparing delta function structures with higher order terms and RHS. 
\paragraph{$ J^\mu_{(2)}$ contribution: }
Since $J^\mu_{(2)}$ is quadratic in $\pi$, we can parameterise it as 
\begin{align}
    J^\mu_{(2)} = \biggl[\mathcal{F}^1_{(2)}\pi \mathcal{F}^2_{(2)}\pi\biggr]^\mu,\label{2ndCurrent}
\end{align}
where $\mathcal{F}^i_{(2)}$  are space time operators $\mathcal{F}^i_{(2)} = \mathcal{F}^i_{(2)}(x^\mu, \partial_\mu)$ which depends on the specific form of off-shell interactions. We denote $G_{(2)}$ as the contribution of $J^\mu_{(2)}$ to the LHS of \eqref{qWT}, which is given by 
\begin{align}
G_{(2)} =&\lim_{q\rightarrow 0} -i q_\mu\int_x e^{iq\cdot x}\prod_{a=1}^n\text{LSZ}_{a+}\braket{0|\mathcal{T}\{J^\mu_{(2)}(x)\pi(x_1)\ldots\pi(x_n)\}|0}\nonumber\\
=&\lim_{q\rightarrow 0} -i q_\mu \int_x  \int_{\vec{p}} \frac{d^d\vec{p}}{2E_p} e^{iq\cdot x} \sum_{L,R} \biggl[\mathcal{F}_{(2)}^1\braket{\{\vec{p}_L\}|\pi(x)|0}\mathcal{F}_{(2)}^2\braket{0|\pi(x)|\Vec{p}} + 1\leftrightarrow 2 \biggr]^\mu\braket{\vec{p}_R,\Vec{p}|0}.
\end{align}
$\{\vec{p}_L\}, \{\vec{p}_R\}$ is a bifurcation of $\{\Vec{p}_1, \Vec{p}_2,\ldots,\Vec{p}_n\}$ and we sum over all possible partitions. One can think of the factorisation as gluing topology depicted in \cite{Cohen:2023ekv} that holds for all loop orders. The matrix element of the overlap between a momentum eigenstate with a field eigenstate is $\braket{\vec{p}_L|\pi(x)|0} = f(\vec{p}_L) e^{i(E_Lt - \vec{p}_L\cdot\vec{x})}$ by space-time translation, where $E_L = \sum_{a \in L} E_a,\; \vec{p}_L = \sum_{a \in L} \vec{p}_a$. Therefore the future time integral would yield the following pole structure:
\begin{align}
  \text{pole structure} = \frac{1}{E_{\vec{p}_L +\vec{q}}(q^0 - E_{\vec{p}_L +\vec{q}} + E_L + i\varepsilon)},  
\end{align}
where $E_L = \sum_{a \in L} E_a,\; \vec{p}_L = \sum_{a\in L} \vec{p}_a,\; E_{\vec{p}_L} = |\vec{p}_L|$. Therefore the far future part in $G_{(2)}$ is,
\begin{align}
    G^+_{(2)} =\lim_{q\rightarrow 0} -i q_\mu  \sum_{L,R}\frac{1}{E_{\vec{p}_L +\vec{q}}(q^0 - E_{\vec{p}_L +\vec{q}} + E_L + i\varepsilon)} \times \biggl[\ldots\biggr]^\mu,
\end{align}
here the information of $\mathcal{F}^i_{(2)}$ are packed into $\biggl[\ldots\biggr]^\mu$ which depends on the concrete model but regular in $q\rightarrow 0$ (since the divergence only comes from future time integral)  with a \textit{soft hierarchy} 
\begin{align}
    q\!<\!<\varepsilon.
\end{align}
The leading order divergence in the soft limit of $G_{(2)}^+$ arises when $L$ only contains a one particle state. The soft pole structure is then
\begin{align}
   G^+_{(2)} =& \lim_{q\rightarrow 0} -i q_\mu \sum_{a=1}^n\frac{1}{E_{\vec{p}_a + \vec{q}}(q^0 - E_{\vec{p}_a + \vec{q}} + E_{\vec{p}_a} + i\varepsilon)}  \times \biggl[\ldots\biggr]^\mu\\
    =& \lim_{q\rightarrow 0} -i q_\mu \sum_{a=1}^n \frac{1}{E_{\vec{p}_a +\vec{q}}(q^0 - \vec{q}\cdot\frac{\vec{p}_a}{E_a} + i\varepsilon)}  \times \biggl[\ldots\biggr]^\mu.
\end{align}
$G^+_{(2)}$ can be further divided into terms that are regular and divergent in the colinear limit $q^0 E_a - \vec{q}\cdot\vec{p}_a \rightarrow 0$ respectively
\begin{align}
    G^+_{(2)} = G^+_{(2)}|_{reg} + G^+_{(2)}|_{div}.
\end{align}
The divergent term $G^+_{(2)}$ upon imposing soft hierarchy, vanishes in the soft limit by sending the front factor $q_\mu \rightarrow 0$ as we keep  $i\varepsilon$ fixed. Moreover, the regular term is completely determined by the free field equation of motion and symmetry variation, we will explicitly demonstrate this statement in the context of EFT of inflation in Section \ref{Sec5}. Here we could see the significance of the soft hierarchy, if we do not send $q\rightarrow 0$ before $\varepsilon\rightarrow 0$, the front factor $q_\mu$ is not soft enough to cancel out the pole in $G^+_{(2)}|_{div}$, which would yield an indeterminate limit. However, if we impose the soft hierarchy where we treat $\varepsilon$ as an IR regulator, $G^+_{(2)}|_{div}$ again vanishes. Same applies to the intermediate and far past integral, yielding
\begin{align}
    G_{(2)} = G_{(2)}^+ + G_{(2)}^0 + G_{(2)}^- = G^+_{(2)}|_{reg}.
\end{align}
Upon sending $q_\mu \rightarrow 0$, $G_{(2)}$ is schematically
\begin{align}
    G_{(2)} = G^+_{(2)}|_{reg} = \sum_{a=1}^n\mathcal{O}_L(\vec{p}_a,\partial_{\vec{p}_a})\tilde{\mathcal{A}}^n
\end{align}
where $\mathcal{O}_L(\vec{p}_a, \partial_{\vec{p}_a})$ are determined by the current \eqref{2ndCurrent}, for example, if the linear part of the symmetry is rotation, then
 \begin{align}
     \mathcal{O}_L(\vec{p}_a, \partial_{\vec{p}_a}) = (i) r_{ij}(p_a^i\partial_{p_a}^j-p_a^j\partial_{p_a}^i).
\end{align}

\paragraph{$J^\mu_{(m)}$ contributions:}

The pole structure for $J^\mu_{(m)}$ when $m>2$ is the same as $J^\mu_{(2)}$, since the matrix element $\braket{\{\vec{p}_L\}|\pi(x)|0}$ can always be factorised into $f(\vec{p}_L) e^{i\sum_{a \in L}p_a\cdot x} $, thus yielding the same time integral. As a result, it will always vanish due to the soft momentum factor in front
\begin{align}
    G_{(m)} = 0\;\; \text{for}\;m > 2.
\end{align}
In conclusion, the only contribution to the LHS of \eqref{qWT} arises from the leading and sub-leading order divergence,
\begin{keyeqn}
    \begin{align}
    \text{LHS} = G_{(1)} + G^+_{(2)}|_{reg},
\end{align}
\end{keyeqn}
which solely depends on the free equation of motion and symmetry transformation upon imposing a soft hierarchy
\begin{align}
    q\!<\!< \varepsilon,
\end{align}
so that terms that are either divergent or indeterminate in the soft limit do not enter the soft theorem. Therefore the LHS of the soft theorem has no off-shell interaction dependence that is not constrained by symmetry.
\subsection{RHS of \eqref{qWT}}
Now we focus on the RHS of Ward-Takahashi identity. Non(sub)-linear, quadratic and higher orders in $\pi$ don't generate pole at $p_a^0 = E_a$ which follows the same argument from the LHS. Since the LSZ formalism specifically extracts the residue at the pole, these terms are excluded by the same reasoning applied to the LHS. Therefore the RHS of \eqref{qWT} is then
\begin{align}
   \text{RHS} =&-i\sum_{b=1}^n \lim_{q\rightarrow 0} \int_x  \prod_{a=1}^n \text{LSZ}_{a+}e^{iq\cdot x} \delta^D(x-x_a)  \braket{0|\pi(x_1)\ldots\delta_{(1)}\pi(x_b)\ldots \pi(x_n)|0} \nonumber\\
    =&-i \sum_{b=1}^n \lim_{q\rightarrow 0} \prod_{a=1}^n\text{LSZ}_{(p_a + q)+} \braket{0|\pi(x_1)\ldots\delta_{(1)}\pi(x_b)\ldots \pi(x_n)|0}.
\end{align}
where $\sum\text{LSZ}_{(p_a + q)+}$ implies we insert $p_a+q = (p_a^0 + q^0, \vec{p}_{a}+\vec{q})$ for particle $a$ while other particle insertions remain the same, and we sum over all particle choices $a$ as required by the Ward-Takahashi identity. The schematic pole contribution is then
\begin{align}
   \text{pole structure} = \lim_{q\rightarrow 0}\lim_{p_a^0\rightarrow E_a}\frac{p_a^2}{E_{p_a+q}(p_a^0 + q^0 - E_{p_a+q} )}.
\end{align}
The limit is indeterminate, therefore according to the prescription we apply $i\varepsilon$ shift on the pole located at $p_a^0 = E_{\vec{p}_a+ \vec{q}} - q^0$. Therefore, the RHS yields zero trivially upon sending $p_a^0 \rightarrow E_a$
\begin{keyeqn}
   \begin{align}
    \text{RHS} = 0 \label{linear},
\end{align} 
\end{keyeqn}
and does not enter the soft theorems. The $i\varepsilon$ prescription ensures the on-shell and soft limit commute, yielding an unambiguous result for the limit.
\subsection{Putting everything together}
By equating both sides and comparing the delta function structure with the non-linear contribution coming from LHS of Ward-Takahashi identity, we arrive at the following soft theorem for $\delta_{(0)}\pi$ that is purely spatial 
 $\delta_{(0)}\pi = b_{i_1\ldots i_N}x^{i_1}\ldots x^{i_N}$ (we still allow for field-dependent terms)
\begin{align}
    b_{i_1\ldots i_N} \lim_{\vec{p}\rightarrow 0}\biggl\{  \partial_{p_{i_1}}\dots\partial_{p_{i_N}}\left[\frac{\tilde{\mathcal{A}}^{n+1}_{E_{p}}+\tilde{\mathcal{A}}^{n+1}_{-E_{p}}}{2}\right] \biggr\} & = -\sum_a \mathcal{O}_L(\vec{p}_a, \partial_{\vec{p}_a})\tilde{\mathcal{A}}^n,\\
     b_{i_1\ldots i_k}\lim_{\vec{p}\rightarrow 0} \biggl\{  \partial_{p_{i_1}}\dots\partial_{p_{i_k}}\left[\frac{\tilde{\mathcal{A}}^{n+1}_{E_{p}}+\tilde{\mathcal{A}}^{n+1}_{-E_{p}}}{2}\right] \biggr\} & =0 ,\;\;k=0,1,\ldots, N-1.
\end{align}
The factors of $i$ herein are implicitly absorbed into the RHS. Notice that the quadratic term $G^+_{(2)}|_{reg}$ does not involve derivatives of delta functions which is guaranteed by the explicit space-time independence of the Hamiltonian, therefore it only matches the leading order $G_{(1)}$ when all derivatives hit the scattering amplitude where derivatives hitting delta functions would yield a tower structure \cite{Du:2024hol}. If the non-linear symmetry $\delta_{(0)}\pi$ involves time $t$, i.e. $\delta_{(0)}\pi =c_{i_1 i_2 \ldots i_N} t x^{i_1} x^{i_2}\ldots x^{i_N}$, then the soft theorem is
\begin{align}
      c_{i_1\ldots i_N}\lim_{\vec{p}\rightarrow 0} \biggl\{ \partial_{p_{i_1}}\dots\partial_{p_{i_N}}\left[\frac{\tilde{\mathcal{A}}^{n+1}_{E_{p}}-\tilde{\mathcal{A}}^{n+1}_{-E_{p}}}{2E_{p}}\right] \biggr\} & = -\sum_a \mathcal{O}_{L}(\vec{p}_a, \partial_{\vec{p}_a})\tilde{\mathcal{A}}^n ,\\
      c_{i_1\ldots i_k}\lim_{\vec{p}\rightarrow 0} \biggl\{ \partial_{p_{i_1}}\dots\partial_{p_{i_k}}\left[\frac{\tilde{\mathcal{A}}^{n+1}_{E_{p}}-\tilde{\mathcal{A}}^{n+1}_{-E_{p}}}{2E_{p}}\right] \biggr\} & = 0,\;\;k=0,1,\ldots N-1\\ 
     c_{i_1\ldots i_k}\lim_{\vec{p}\rightarrow 0} \biggl\{  \partial_{p_{i_1}}\dots\partial_{p_{i_k}}\left[\frac{\tilde{\mathcal{A}}^{n+1}_{E_{p}}+\tilde{\mathcal{A}}^{n+1}_{-E_{p}}}{2}\right] \biggr\} & = 0,\;\;k=0,1,\ldots N. 
\end{align}
We imposed the energy momentum conserving deltas on $\mathcal{A}^n$ to promote it to $\tilde{\mathcal{A}}^n$ before acted upon $\mathcal{O}_L(\vec{p}_a, \partial_{\vec{p}_a})$ such that the delta dependence is completely eliminated and the equations only depend on on-shell amplitudes. We see the emergence of the tower structure which is entailed by the closure of algebra with space-time translations as explained in \cite{Du:2024hol}. These two possibilities exhaust all options because the requirement for the free theory to be a two-derivative theory does not allow for higher orders of $t$ in $\delta_{(0)}\pi$. 
\subsection{Soft hierarchy} 
From the above derivation, the soft theorem comes from amputating the Ward-Takahashi identity
\begin{align}
    &\lim_{q\rightarrow 0}\int_{x} e^{iq\cdot x}\prod_{a=1}^n\text{LSZ}_{a^+}\partial_{\mu} \braket{0|\biggl[J^\mu_{(1)}(x) + J^\mu_{(2)}(x) \biggr]\pi(x_1) \ldots \pi(x_n)|0} 
    = 0,
\end{align}
where the result only cares about current and symmetry transformation that are linear in $\pi$ since the quadratic current that contributes to $G^+_{(2)}$ is determined by the symmetry variation and free EOM. The soft limit and on-shell limit from LSZ trivialises the non-linear contribution as we showed above. The formula establishes a clear connection between the amplitudes of soft Nambu-Goldstone mode emissions and the amplitudes of hard processes. Implicitly, we imposed a soft scale hierarchy  
\begin{align}
    q \sim p_a^0 - E_a <\!\!< \varepsilon<\!\!<p_a,
\end{align}
i.e. we send $q\rightarrow 0$ and $p_a^0 - E_a\rightarrow 0$ while keeping $\varepsilon$ fixed, so that $i\varepsilon$ regulates all IR divergence in the soft limit, thus making the soft theorem innately valid to all-order perturbations that does not depend on any off-shell interaction that is not constrained by symmetry including cubic vertex.  Again, it's worth stressing that only the indeterminate and divergent terms receive $i\varepsilon$ shift, while terms that are regular in the limit under consideration remain unchanged to ensure that the amplitude remains field redefinition invariant. To illustrate, consider a trivial theory up to field redefinition
\begin{align}
    S = \int d^D x\;\; \frac{1}{2}\partial_\mu\phi\partial^\mu\phi\bigl(1+ \lambda\phi\bigr).
\end{align}
The tree level four point amplitude is then 
\begin{align}
    \mathcal{A}_4 \sim \frac{s_{1,2}^2}{s_{1,2}+i\varepsilon E_{12}} + \frac{s_{1,3}^2}{s_{1,3}+i\varepsilon E_{13}} + \frac{s_{1,4}^2}{s_{1,4} + i\varepsilon E_{14}}.
\end{align}
If we explicitly include the $i\varepsilon$ and send one momentum soft, then $\mathcal{A}_4$ would have non-zero soft scaling:
\begin{align}
    \mathcal{A}_4 \sim \mathcal{O}(p_1^2),
\end{align}
which would break field redefinition invariance. Therefore, the only consistent prescription one should take is to only include $i\varepsilon$ when the soft limit is divergent or indeterminate. In the above example, the amplitude in the soft limit is clearly regular and determinate
\begin{align}
    \mathcal{A}_4 \sim s_{1,2} + s_{1,3} + s_{1,4} = 0.
\end{align}
Hence the $i\varepsilon$ could be safely discarded.
Another key observation is that the soft amplitude combination $\tilde{\mathcal{A}}_{E_p} + \tilde{\mathcal{A}}_{-E_p}$ could never cancel with any soft function $S(\vec{p})$ that contains a soft pole due to the presence of cubic vertices
\begin{align}
    \lim_{\vec
    {p}\rightarrow 0}\tilde{\mathcal{A}}^{n+1}_{E_p} + \tilde{\mathcal{A}}^{n+1}_{-E_p} &\supset \frac{(\ldots)}{E_p E_a - \vec{p}\cdot\vec
    {p}_a} + \frac{(\ldots)}{-E_p E_a - \vec{p}\cdot\vec
    {p}_a}, \\
    \lim_{\vec
    {p}\rightarrow 0}S(\vec
    p)\tilde{\mathcal{A}}^n&\supset \frac{(\ldots)}{E_p E_a - \vec{p}\cdot\vec
    {p}_a}
\end{align}
The analytic structure in the soft limit for both scenarios are essentially different due to the combination with the soft energy flip unless the pole is regulated by $i\varepsilon$.
To summarise, there are two benefits of the soft hierarchy
\begin{itemize}
    \item Off-shell interaction including cubic vertices that are not constrained by symmetries are completely eliminated by the front factor $q_\mu$ from Ward-Takahashi identity as $\varepsilon$ regulates all IR divergence.
    \item With the soft hierarchy, any inderterminate limits including soft limit and on-shell limit automatically commutes, resulting in no ambiguity with respect to the order of the two limits.
\end{itemize}
Besides these two benefits, we would also like to argue that the soft hierarchy captures the \textit{true} IR behaviour of the theory. The $S$-matrix, on its own, is not directly observable—the actual measurable quantity is the cross-section. To derive the cross-section from the $S$-matrix, it must be integrated over the phase space. However, the $S$-matrix behaves as a distribution, meaning it can diverge at certain points within the phase space. Regulating this integral requires introducing $\varepsilon$ as an IR regulator in the integrand. As we perform the phase space integration, $\varepsilon$ remains fixed, with the limit 
$\varepsilon\rightarrow 0$ applied only at the end. This process ensures that as we traverse the phase space, we inevitably reach points where the soft hierarchy holds, specifically where the soft momenta are significantly smaller than $\varepsilon$.

\section{Hidden Adler zeros for $\dot\pi^3$ }\label{Sec4}

For theories with shift symmetries $\pi \rightarrow \pi +c$, Adler zero is regarded as a fundamental property of the scattering amplitude, i.e. the scattering amplitude vanishes if we send one momentum to zero. However, in the presence of a cubic vertex, as can arise when we break Lorentz boosts or when we couple to additional degrees of freedom, the conclusion seems to break down since the propagator $\frac{1}{(p + p_a)^2}= \frac{1}{2E E_a - 2\vec{p}\cdot\vec{p}_a }$ is divergent in the soft limit $\vec{p}\rightarrow 0$. This is also clear in the derivation, the factor $q^\mu$ is not soft enough to annihilate the soft pole arises from soft cubic vertices. However, from the derivation, we see that the $i\varepsilon$ is introduced to regulate the far past/future time integral. If we demand that the soft momentum vanishes faster than $i\varepsilon$ as dictated by our soft hierarchy, the contribution is again zero, thus making Adler zero manifest at all order perturbations. 

In this section we present a simple example to illustrate how the hidden Adler zeros work in practice. In the previous sections we denoted the soft momentum as $p^\mu = (E_p, \vec{p})$, however for the sake of convenience we usually take $p_1$ as the soft momentum when doing explicit calculations, henceforth we identify $p_1^\mu = p^\mu$ which means we take $p_1$ soft.
Consider the cubic vertex $\dot\pi^3$ which comes from the EFT of inflation, which is clearly invariant under all spatial shift $\delta\pi = b_{i_1\ldots i_N}x^{i_1}\ldots x^{i_N}$, thus any amplitude coming from the vertex should satisfy all spatial soft theorems. The 3-point amplitude is $\tilde{\mathcal{A}}^3_{E_1} \sim E_1 E_2 (E_1 + E_2) $, where we've eliminated $E_3$ by going to the minimal basis. Then the amplitude combination is  $\tilde{\mathcal{A}}^3_{E_1} + \tilde{\mathcal{A}}^3_{-E_1} \sim E_1^2 E_2 $, hence the spatial soft theorems for such symmetry \cite{Du:2024hol}
\begin{align}
   \lim_{\vec{p}\rightarrow 0} b_{i_1\ldots i_k} \partial_{p_{i_1}}\ldots\partial_{p_{i_k}}( \tilde{\mathcal{A}}_{E_p} + \tilde{\mathcal{A}}_{-E_p}) = 0, \;\;\;k = 0, 1, 2, \ldots
\end{align}
are satisfied to all orders if $b_{ij}$ is traceless, since $\partial_{p_i}\partial_{p_j}E^2= \delta^{ij} $, and the invariance of the free theory requires these parameters $b_{i_1\ldots i_N}$ to be traceless. At 4-point, the $s$ channel amplitude is
\begin{align}
    \tilde{\mathcal{A}}^4_{E_1} \supset \frac{E_1 E_2 (E_{1} + E_2)^2 E_3 (E_1 + E_2 + E_3)}{s_{1,2} +i\varepsilon E_{12}},
\end{align}
where $E_{ab}=|\vec{p}_a+\Vec{p}_b|$. Taylor expanding the amplitude around $\Vec{p}_1/\varepsilon = 0$, schematically we get
\begin{align}
    \tilde{\mathcal{A}}^4_{E_1} \supset \frac{E_1}{i\varepsilon} E_2^3 E_3 (E_2 + E_3) + \frac{E_{1}}{i\varepsilon}\biggl[\mathcal{O}\biggl(\frac{\Vec{p}_1\cdot\Vec{p}_c}{i\varepsilon}\biggr) + \mathcal{O}\biggl(\frac{E_1}{i\varepsilon}\biggr)\biggr]\times \text{hard momentum structure}.
\end{align}
The amplitude combination that appears in the soft theorem  is then
\begin{align}
    \lim_{\vec{p}\rightarrow 0}\tilde{\mathcal{A}}^4_{E_1} + \tilde{\mathcal{A}}^4_{-E_1} \supset \mathcal{O}\left[\biggl(\frac{E_1}{i\varepsilon}\biggr)^2\right] .
\end{align}
Hence in the soft limit, the momentum derivatives of the amplitude combination vanish while contracting with traceless parameters.  Notice the soft hierarchy $E_1 <\!\!< \varepsilon$ is crucial for the soft theorem to hold which is consistent with the soft theorem derivation from Section \ref{Sec3}. From the soft theorem derivation, we've observed the emergent Adler zero \eqref{AdlerZero}, and the amplitude from the vertex indeed satisfies the Adler zero condition.

\section{EFT of inflation} \label{Sec5}
In the context of EFT of a single field inflation, time diffeomorphism is broken by the Friedmann –Lemaitre – Robertson – Walker (FLRW) ground state solutions. The fluctuation $\pi$ around the background non-linearly realises time diffeomorphism. In the flat space and decoupling limit, the non-linear time diffeo reduces to non-linear boost symmetry and therefore connects with what we are considering in this work. We refer the readers to \cite{Grall:2020ibl, Green:2022slj,Creminelli:2022onn} for a more comprehensive discussion on the flat space and decoupling limit of inflation. Moreover, the flat space scattering amplitude appears as the residue of the $k_T$ pole of the inflationary wavefunction coefficient \cite{Raju:2012zr,Maldacena:2011nz}. Thus, studying flat space scattering amplitude is a crucial and necessary step towards understanding the wavefunction coefficients and the cosmological correlators. 

In this section we will discuss the soft theorem for non-linearly realised Lorentz boost that arises from the flat space and decoupling limit of EFT of inflation and show that the soft theorem is powerful enough to uniquely determine the Lagrangian up to known degrees of freedom for the Wilson coefficients.

In the above limits, at leading order the Lagarangian is effectively described by a superfluid EFT. The superfluid Lagrangian is given by expanding a Lorentz invariant one derivative per field theory by a single clock background $\braket{\phi} = t$\footnote{We set the symmetry breaking scale or the chemical potential to $1$}
\begin{align}
    \mathcal{L} = X + M_2 X^2 + M_3 X^3 + M_4 X^4 + M_5 X^5 + \ldots
\end{align}
where $X = \frac{1}{2} (\partial_\mu \phi \partial^\mu \phi - 1) = \dot\pi + \frac{1}{2}\dot\pi^2 - \frac{1}{2}(\partial_i\pi)^2 $. The parameter $M_2$ can be substituted by the 'sound speed'\footnote{Since we have canonically normalised $\pi$, the 'sound speed' is only considered as a parameter of the theory but not the propagation speed for the field.} via $M_2 = \frac{1}{c_s^2} - 1$. Under a single clock background $\braket{\phi} = t$, Lorentz boosts are non-linearly realised as
\begin{align}
    \delta_B \pi = b_i[x^i + (x^i\partial_t + t\partial_{i})\pi].
\end{align}
The quadratic current upon canonical normalisation \footnote{ When the sound speed is less than one, we need to perform a coordinates and field rescaling such that the free theory remains proportional to $\frac{1}{2}(\partial_\mu\pi)^2$
\begin{align}
    t\rightarrow t/c_s.
\end{align}
The soft theorem also rescales accordingly. Moreover, we rescale the field by
\begin{align}
  \pi \rightarrow \sqrt{c_s}\pi,  
\end{align}
such that the kinematic term is properly normalised.
} is then
\begin{align}
    &\partial_\mu J^\mu_{(2)} \nonumber\\
    =&b_i\biggl\{\underbrace{\partial_\mu \biggl[\frac{1}{\sqrt{c_s}}\partial^\mu \pi \biggl(x^i\partial_t +  t\partial_i \biggr)\pi\biggr]}_\text{regular} + \underbrace{3 M_3 \partial_t ( x^i \dot\pi^2) -\frac{1}{2\sqrt{c_s}}\biggl[\partial_t\biggl(x^i\partial_\mu\pi\partial^\mu\pi\biggr) + \partial_i\biggl(t\partial_\mu\pi\partial^\mu\pi\biggr)  \biggr]}_{\text{divergent}}\biggr\}.
\end{align}
Therefore only the first term enters the $G^+_{(2)}|_{reg}$,
\begin{align}
    G^+_{(2)}|_{reg} = &\lim_{q\rightarrow 0} \frac{-i}{\sqrt{c_s}} q_\mu \int_{t>T^+} \int_{\vec{x}}  \int_{\vec{p}} \frac{d^d\vec{p}}{2E_p} e^{iq\cdot x} \sum_{a=1}^n \braket{\vec{p}_1,\ldots,\Vec{p},\ldots\vec{p}_n|0}_a \nonumber\\
    &\biggl[\braket{\vec{p}_a|\partial^\mu\pi(x)|0}\braket{0|(x^i\partial_t + t\partial_i)\pi(x)|\Vec{p}} + \braket{0|\partial^\mu\pi(x)|\vec{p}}\braket{\vec{p}_a|(x^i\partial_t + t\partial_i)\pi(x)|0}\biggr].
\end{align}
Here $\braket{\vec{p}_1,\ldots,\Vec{p},\ldots\vec{p}_n|0}_a$ highlights the replacement of $\vec{p}_a$ with $\vec{p}$ within the $S$-matrix element $\braket{\alpha|0}$. We then arrive at
\begin{align}
    G^+_{(2)}|_{reg} = \frac{i}{\sqrt{c_s}} \sum_a E_a \partial_{p_a^i} \braket{\alpha|0}.
\end{align}
The naively divergent terms cancel each other between $x^i\partial_t$ and $t\partial_i$ similar to the cancellation in \eqref{LIBoost}. Notice the factor of $2$ in the denominator is canceled by considering permutation between $p$ and $p_a$ resulting in a single compact term.  By assembling our results from Section \ref{Sec3}, the soft theorems for Nambu-Goldstone mode arising from spontaneous Lorentz boost breaking are:
\begin{keyeqn}
\begin{align}
    \lim_{\vec{p}\rightarrow{}0}  \biggl\{ \partial_{p^i} \left [\frac{\tilde{\mathcal{A}}^{n+1}_{E_{p}} + \tilde{\mathcal{A}}^{n+1}_{-E_{p}}}{2}\right]   \biggr\}&= -i \frac{1}{\sqrt{c_s}}\left[\sum_a E_a\partial_{p_a^i}\tilde{\mathcal{A}}^n\right] \,, \label{EFToI}\\
\lim_{\vec{p} \rightarrow 0} \biggl\{\tilde{\mathcal{A}}^{n+1}_{E_{p}} + \tilde{\mathcal{A}}^{n+1}_{-E_{p}} \biggr\} &= 0\,. \label{adler0}
\end{align}
\end{keyeqn}
 In the flat space and decoupling limit of the EFT of inflation, the broken time diffeomorphism reduces to the breaking of Lorentz boost \cite{Cheung:2007st,Baumann:2011su,Green:2022slj}, therefore the soft theorem we get from the non-linearly realised boost is precisely \eqref{EFToI}. Moreover we see the emergence of Adler zero as a consequence of acting momentum derivatives on the momentum delta function, which is expected from the symmetry algebra which dictates that the non-linear realisation of boosts must also come with a shift symmetry due to the commutator: \cite{Nicolis:2015sra, Roest:2019dxy}
\begin{align}
    [P_i, K_j] = \delta_{ij} (Q - P_0),\label{InverseHiggs}
\end{align}
where $P_{\mu}$ is the space time translation generator, $K_i$ is the non-linear boost, $Q$ is the $U(1)$ generator such that there exists a unbroken combination of time translation and constant shift. The coset for such symmetry breaking pattern is then
\begin{align}
ISO(1,3)\times U(1)/SO(3),
\end{align}
and the coset parameter is given by
\begin{align}
    g = e^{x^\mu P_\mu} e^{\xi^i K_i} e^{\pi Q}.
\end{align}
The Goldstone mode $\xi_i$ can be removed by the inverse Higgs constraint using \eqref{InverseHiggs}. One can find a more detailed discussion on the coset construction for EFT of inflation in \cite{Finelli:2018upr}. We see the nice connection between amplitude soft theorem and symmetry algebra: the inverse Higgs constraints from the algebra are translated to the tower structure of the amplitude soft theorem \cite{Du:2024hol}. The soft theorems we have derived in this work can be tested against the superfluid amplitudes, but can also be used to bootstrap the amplitudes by starting with an ansatz that is rotationally invariant. Indeed, the soft theorem is powerful enough to fix the Wilson coefficients of the amplitude ansatz up to known degrees of freedom. We will show this explicitly below before first illustrating how the soft theorem works in practice. 
\subsection{Relativistic example}
We provide a simple relativistic example to demonstrate how the soft theorem works in practice. The interaction Lagrangian density for a generic Lorentz boost breaking theory with one derivative per field is parameterised as
\begin{align}
    \mathcal{L}_I &= g_{3,3} \dot\pi^3 + g_{3,1} \dot\pi (\partial_i\pi)^2 \nonumber\\
    &+ g_{4,4} \dot\pi^4 + g_{4,2}\dot\pi^2(\partial_i\pi)^2 + g_{4,0}(\partial_i\pi)^4 \nonumber\\
    &+ g_{5,5} \dot\pi^5 + g_{5,3} \dot\pi^3 (\partial_i\pi)^2 + g_{5,1}\dot{\pi}(\partial_i\pi)^4,
\end{align}
where $g_{m,n}$s are the Lagrangian Wilson coefficients. We set the 'sound speed' $c_s$ to $1$ and write down an amplitude ansatz \cite{Green:2022slj,Cheung:2023qwn}
\begin{align}
    \tilde{\mathcal{A}}^3 &= -i6g_3 E_1 E_2 (-E_1 - E_2), \\
    \tilde{\mathcal{A}}^4_{E_1} &= 24 g_{4,4} E_1 E_2 E_3 (- E_1 - E_2 - E_3 ) + g_{4,2} \sum_{\text{perms}}E_{1}E_{2} (\vec{p}_3\cdot\vec{p}_4)|_{\text{minimal basis}} \nonumber\\
    &+ g_{4,0}\sum_{\text{perms}} (\vec{p}_1\cdot\vec{p}_2)(\vec{p}_3\cdot\vec{p}_4)|_{\text{minimal basis}} + 18  g_3^2\sum_{\text{perms}} \frac{E_1 E_2 (E_{1} + E_2)^2 E_3 E_4}{s_{1,2} +i\varepsilon E_{12}}|_{\text{minimal basis}}.
\end{align}
The scattering amplitude is computed as follows: First, we evaluate the amplitude in the minimal basis without applying the $i\varepsilon$ prescription. Then, the appropriate $i\varepsilon$ term is introduced to the soft co-linear pole, modifying $\frac{1}{s_{1,a}}$ to $\frac{1}{s_{1,a} + i\varepsilon E_{1a}}$ to serve as an IR regulator.
The kinematic dependence at coupling order $g_3^2$ implies we set the Wilson coefficient $g_{3,1}$ of the vertex $\dot\pi(\partial_i\pi)^2$ to zero. 
We first examine the soft theorem at $n=3$ which relates four-point soft amplitude with three-point hard ones. The amplitude combination under minimal basis is then
\begin{align}
    &\left[\frac{\tilde{\mathcal{A}}_{E_1}^4 + \tilde{\mathcal{A}}_{-E_1}^4 }{2}\right] \nonumber\\
    = &-4 g_{4,2} \biggl\{ E_1^2\left[2(E_2^2 + E_3^2 + 3E_2 E_3) - \vec{p}_1\cdot\vec{p}_2 -\vec{p}_1\cdot\vec{p}_3 \right] + \vec{p}_1\cdot\vec{p}_2 (E_3^2 + 2E_2E_3) + \vec{p}_1\cdot\vec{p}_3 (E_2^2 + 2E_2E_3) \biggr\}\nonumber\\
    &-8g_{4,0}\biggl\{-2(\vec{p}_1\cdot\vec{p}_2)^2 -2(\vec{p}_1\cdot\vec{p}_3)^2 + E_1^2(E_2 E_3 -\vec{p}_1\cdot\vec{p}_2 -\vec{p}_1\cdot\vec{p}_3) \nonumber\\
    &+\vec{p}_1\cdot\vec{p}_2 (E_3^2 + 2E_2E_3) + \vec{p}_1\cdot\vec{p}_3 (E_2^2 + 2E_2E_3)  \biggr\}  \nonumber\\
    & +  g_3^2 \mathcal{O}\left[\biggl(\frac{E_1}{i\varepsilon}\biggr)^2\right] + 24 g_{4,4} E_1^2 E_2 E_3.
\end{align}
Here, the term $\vec{p}_1 \cdot\vec{p}_a$ arises from the amplitude combination since
\begin{align}
    \frac{s_{1,a}(E_1) + s_{1,a}(-E_1)}{2} =-\vec{p}_1\cdot \vec{p}_{a} .
\end{align}
The LHS of \eqref{EFToI} reads
\begin{align}
    \lim_{\vec{p}_1\rightarrow 0 }\partial_{\vec{p}_1}\left[\frac{\tilde{\mathcal{A}}^4_{E_1} + \tilde{\mathcal{A}}^4_{-E_1}}{2}\right] = -(4g_{4,2} + 8g_{4,0})\left[\vec{p}_2(E_3^2 + 2E_2E_3) + \vec{p}_3 (E_2^2 + 2E_2E_3)\right].
\end{align}
For the RHS we sum over all the hard modes $a=2,3,4$ and find
\begin{align}
    -i\sum_{a=1}^3 E_a\partial_{\vec{p}_a}\tilde{\mathcal{A}}^3 = &6i^2g_3\biggl[(E_2\partial_{\vec{p}_2} + E_3\partial_{\vec{p}_3} + E_4\partial_{\vec{p}_4})[E_{2}E_{3}(-E_{2}-E_{3})]\biggr]\nonumber \\
    =& 6g_3 \left[\vec{p}_2(E_3^2 + 2E_2E_3) + \vec{p}_3 (E_2^2 + 2E_2E_3)\right].
\end{align}
There is no $\vec{p}_4$ dependence in the amplitude as we've already imposed the minimal basis. Equating both sides we get
\begin{align}
    g_{4,2} + 2g_{4,0} = -\frac{3}{2} g_3. \label{EFToIConstraint}
\end{align}
Here we see how the soft theorem is able to relate the Wilson coefficients at different orders in the field, which is exactly what the symmetry enforces given that it contains both zeroth-order and linear-order terms in the field.
However, from the Lagrangian formalism of EFT of inflation, the non-linear symmetry imposes two constraints on the 4-point function where at $n=3$ we only see one constraint. To see how the other constraint emerges, we need to go to higher point. Consider the same theory with $n=4$, the soft theorem then relates the soft five-point amplitude to the hard four-point. Focusing on the $123-45$ and $23-145$ factorisation channel, the five point amplitude is messy to write down, so we only show the terms that contribute to the soft theorem. At the coupling order $g_{4,2}g_3$, the LHS is
\begin{align}
    &\lim_{\vec{p}_1\rightarrow 0}\partial_{\vec{p}_1}\left[\frac{\tilde{\mathcal{A}}^5_{E_1}(123-45)+\tilde{\mathcal{A}}^5_{-E_1}(123-45)}{2} + (23\leftrightarrow 45)\right]\nonumber\\
    \supset& -6 g_{4,2}g_3 \frac{E_2 + E_3}{s_{23}}\biggl\{ \vec{p}_4 E_2 E_3 ( E_2 +E_3 ) (E_2 + E_3 + 2 E_4) \nonumber\\
    +& E_4\biggl[ \vec{p}_2 E_3 \biggl( E_3^2 + 5E_2 E_3 + 4E_2^2 + E_4(3E_2 + E_3) \biggr) +\vec{p}_3 E_2 \biggl( E_2^2 + 5E_2 E_3 + 4E_3^2 + E_4(3E_3 + E_2) \biggr) \biggr]   \biggr\}.
\end{align}
To compute the RHS we write down the expression for the four-point amplitude at coupling order $g_3^2$ and factorisation channel $23-45$
\begin{align}
    \tilde{\mathcal{A}}^4(23-45) = -9 i g_3^2 \frac{E_2 E_3 (E_2 + E_3)^2 E_4 (- E_2 - E_3 - E_4)}{s_{2,3}}.
\end{align}
The RHS of \eqref{EFToI} is then
\begin{align}
    &-i(E_2\partial_{\vec{p}_2} + E_3\partial_{\vec{p}_3} + E_4\partial_{\vec{p}_4} + E_5\partial_{\vec{p}_5}) \tilde{\mathcal{A}}^4(23-45) \nonumber\\
    =& 9 g_3^2 \frac{E_2 + E_3}{s_{23}}\biggl\{ \vec{p}_4 E_2 E_3 ( E_2 +E_3 ) (E_2 + E_3 + 2 E_4) \nonumber\\
    +& E_4\biggl[ \vec{p}_2 E_3 \biggl( E_3^2 + 5E_2 E_3 + 4E_2^2 + E_4(3E_2 + E_3) \biggr) +\vec{p}_3 E_2 \biggl( E_2^2 + 5E_2 E_3 + 4E_3^2 + E_4(3E_3 + E_2) \biggr) \biggr]   \biggr\}.
\end{align}
Notice the propagator is Lorentz invariant, therefore it yields zero when acted upon by $\sum_a E_a\partial_{\vec{p}_a}$. We see the kinematic dependence precisely matches the LHS. Now if we evaluate the LHS at coupling order $g_{4,0}g_3$, the kinematic dependence is essentially different. Thus the soft theorem uniquely picks out the coefficient 
\begin{align}
    g_{4,2} = -\frac{3}{2}g_3. \label{Relativistic4-point}
\end{align}
which further implies $g_{4,0} = 0$ by \eqref{EFToIConstraint}. The result \eqref{Relativistic4-point} completely agrees with Lagrangian analysis, the Wilson coefficients $g_{4,0}$ is proportional to the Wilson coefficient of $\dot\pi(\partial_i\pi)^2$ which we set to zero as an input for this subsection. 
In this example, the soft theorem is only satisfied when summing over all possible soft modes emission for a given hard process, where in our case the hard process is $23-45$, hence we need to consider contributions from both $123-45$ and $23-145$.
\subsection{Bootstrap} 
Now we would like to bootstrap the theory up to 5-point coefficients with arbitrary 'sound speed' $c_s$. First, at the level of $n=3$ the soft theorem constraint \eqref{EFToIConstraint} scales to
\begin{align}
    g_{4,2} + 2 g_{4,0} + 2(g_{3,1})^2  = -\frac{3}{2\sqrt{c_s}} (g_{3,3}+ g_{3,1})  \label{Nonrel4}
\end{align}
where $2(g_{3,1})^2$ arises from the exchange diagram. In the context of EFToI or superfluid, the Hamiltonian Wilson coefficients are given by
\begin{align}
    &g^H_{3,3} = -\sqrt{c_s^{7}} \biggl(\frac{c_s^2-1}{2c_s^2} + M_3\biggr),\;\;g^H_{3,1} = -\frac{1}{2\sqrt{c_s}}(c_s^2 - 1),\nonumber\\
    &g^H_{4,4} = \frac{1}{8} c_s^3 \biggl[8 + 9c_s^4(1 - 2M_3)^2 + c_s^2(-8M_4 + 24M_3 -17)\biggr],\;\;g^H_{4,2} = \frac{c_s}{4} \biggl[c_s^4(6M_3 - 3) + 5c_s^2 -2\biggr],\nonumber\\
    &g^H_{4,0} = \frac{c_s(c_s^2 -1)}{8},\nonumber\\
    &g^H_{5,5} = -\frac{\sqrt{c_s^9}}{8}\biggl\{21 + c_s^2\biggl[-69 + 84M_3 + 3c_s^2(2M_3 -1)(-25 + 9c_s^2(2M_3-1)^2 + 30M_3 -16M_4) - 32M_4 + 8M_5\biggr] \biggr\},\nonumber\\
    &g^H_{5,3} = \frac{c_s^3}{8}\biggl[8 + 9c_s^4(2M_3 - 1)^2 + c_s^2(-17 + 24M_3 -8M_4)\biggr],\;\;g^H_{5,1} = \frac{\sqrt{c_s}}{8}\biggl[-1 + 5c_s^2 -7 c_s^4 + c_s^6(3 - 6M_3)\biggr],
\end{align}
where the Hamiltonian is parameterised by
\begin{align}
    \mathcal{H}_I &= g_{3,3} \tilde\pi^3 + g_{3,1} \tilde\pi (\partial_i\pi)^2 \nonumber\\
    &+ g_{4,4} \tilde\pi^4 + g_{4,2}\tilde{\pi}^2(\partial_i\pi)^2 + g_{4,0}(\partial_i\pi)^4 \nonumber\\
    &+ g_{5,5} \tilde\pi^5 + g_{5,3} \tilde\pi^3 (\partial_i\pi)^2 + g_{5,1}\tilde{\pi}(\partial_i\pi)^4.
\end{align}
Here $\tilde{\pi}$ is the conjugate momentum of the inflaton fluctuation $\pi$. We put the derivation details in Appendix \ref{LagHam}. The above constraint \eqref{Nonrel4} can be recast as
\begin{align}
    g_{4,2}^H +2 g_{4,0}^H = -\frac{3}{2} \sqrt{c_s}^3 (g_{3,3}^H + g_{3,1}^H),
\end{align}
given $g_{3,1}$ as an input. Now we would like to check if the solutions to the soft theorems match the Hamiltonian Wilson coefficients which obey non-trivial relations as dictated by the non-linear realisation of boost symmetry. From the previous section, we've learned that to uncover the additional constraint on 4-point coefficients, we need to examine the 5-point case. First, we disregard any diagram where the soft leg is associated with a cubic vertex featuring a collinear pole located at located at $s_{1,a} = 0$ under minimal basis, and we can show it vanishes upon summing over all channels which will also be discussed in the next section. Then by matching the kinematic dependence on both sides of the soft theorem, we're able to conclude the following constraints
\begin{align}
    g^H_{4,0} &= -\frac{1}{4}\sqrt{c_s}^3 g^H_{3,1},\\
    g^H_{5,3} &= \sqrt{c_s}^3(-g^H_{4,2} - 2g^H_{4,4}) + 12 g^H_{3,3}g^H_{4,0},\\
    g^H_{5,1} &=\frac{1}{2}\sqrt{c_s}^3 (-g^H_{4,2} - 2g^H_{4,0}) + 2 g^H_{3,1}g^H_{4,0}.
\end{align}
The amplitude can be divided into two pieces, one is regular in co-linear pole located at $s_{2,3}=0,\;s_{2,4}=0,\;s_{2,3} + s_{2,4} = 0$; the other is divergent. The regular terms come from contact diagrams and exchange diagrams that are effectively contact, i.e. the propagator is canceled by the vertex. The divergent terms only come from exchange diagrams. The matching between the divergent pieces on both sides of the soft theorem yields the first constraint that only contains 4-point and 3-point coefficients, the matching between regular pieces yields the two latter constraints that determines two 5-point coefficients in terms of lower point coefficients. The matching can be understood diagrammatically via Figure \ref{SFcheck}. Therefore, we obtained two constraints at 4-point and two constraints at 5-point which precisely match the constraints acquired via Hamiltonian analysis. Essentially, the Wilson coefficient of the vertex $\dot{\pi}^n$ is not constrained by the non-linear boost symmetry, therefore there is one unconstrained DoF at each level. Moreover, the constraints from non-linear boost symmetry are notably more structured and straightforward when applied to Hamiltonian coefficients compared to Lagrangian coefficients.

One may ask if the soft theorem is powerful enough to pin down all the Hamiltonian Wilson coefficients up to known DoF. First let's consider the case where $n = 2m+1$ where the soft theorem relates a $2m+2$ pt soft amplitude to a $2m+1$ pt hard amplitude. On the RHS, there are exactly $m+1$ non-degenerate independent $2m+1$ pt hard amplitudes when $m>1$. Since there s an odd number of space-time derivatives in the Lagrangian, the amplitude is not Lorentz invariant, and thus there is no degeneracy when acted upon the boost operator $\sum_a E_a\partial_{\vec{p}_a}$. Consequently, we get $m+1$ independent constraints. On the LHS, there are $m+2$ independent $2m+2$ pt soft amplitudes when $m>1$. Therefore, only one unknown DoF remains, which agrees with the Hamiltonian analysis. The only exception is for $n=3$, since there's only one independent 3-point amplitude. However, from the above analysis we see the additional constraint arises at $n=4$, thus saturating all constraints.

Now for $n=2m+2$, on the RHS, there are $m+2$ independent hard $2m+2$ pt amplitudes. The only Lorentz invariant term is $(\partial_\mu\pi)^{2m+2}$, meaning there would be $m+1$ independent kinematics when acted upon the boost operator, leading to $m+1$ independent constraints from the soft theorem. The LHS has $m+2$ DoF, again the soft theorem saturate all constraints that are imposed by the non-linear symmetry.

In summary, we have checked that the soft theorem for the superfluid i.e. the soft theorem arising due to non-linear boosts, is satisfied by the superfluid amplitudes. It is also powerful enough to in principle bootstrap all tree level amplitudes up to known degrees of freedom given the cubic vertex and this has been checked explicitly up to 5-point amplitude.

It is worth emphasising that the scattering amplitude soft theorems for non-linearly realised boost symmetry cannot fix the coefficients of the cubic vertices. However, as was shown in \cite{Hui:2022dnm}, the unequal-time correlator soft theorem for the same symmetry reveals the relations between cubic interactions and quadratic action. In the flat space limit, we cannot distinguish the two independent cubic vertices $\dot{\pi}^3$ and $\dot{\pi}(\partial_i\pi)^2$ at the level of scattering amplitude since they are degenerate on-shell. Consequently, the information related to the expansion of our universe is obscured by the flat space approximation.
\begin{figure}
    \centering
\begin{align}
&\begin{tikzpicture}[baseline=(m.base)]
    \begin{feynman}
      \vertex[draw,circle,fill=black!20,minimum size=0.05cm] (m) at ( 0, 0) ;
      \vertex (a) at (0.7,0.7) {};
      \vertex (b) at ( 0, 1.5) {};
      \vertex (c) at (-1.29, -0.75) {};
      \vertex (d) at ( 1.29, -0.75) {};
      \diagram* {
      (m) -- [red, fermion] (a),
      (m) -- [fermion] (b),
      (m) -- [fermion] (c),
      (m) -- [fermion] (d),
      };
    \end{feynman}
  \end{tikzpicture}
  +
    \begin{tikzpicture}[baseline=(m.base)]
    \begin{feynman}
      \vertex (m)[draw,circle,fill=black!20,minimum size=0.05cm] at ( 0, 0) ;
      \vertex(n) at ( 1, 0) ;
      \vertex (a) at (-1,-1) {};
      \vertex (b) at ( -0.7, 0.7) {};
      \vertex (c) at (2, 1) {};
      \vertex (d) at ( 2, -1) {};
      \vertex (e) at ( 2, 0) {};
      \diagram* {
      (m) -- [fermion] (a),
      (m) -- [red, fermion] (b),
      (m) -- [dashed] (n),
      (n) -- [fermion] (c),
      (n) -- [fermion] (d),
      };
    \end{feynman}
  \end{tikzpicture}
  +
  \text{perms}
  \rightarrow
  \begin{tikzpicture}[baseline=(m.base)]
    \begin{feynman}
      \vertex[draw,circle,fill=black!20,minimum size=0.05cm] (m) at ( 0, 0) ;
      \vertex (a) at (0.7,0.7) {};
      \vertex (b) at ( 0, 1.5) {};
      \vertex (c) at (-1.29, -0.75) {};
      \vertex (d) at ( 1.29, -0.75) {};
      \diagram* {
      (m) -- [fermion] (b),
      (m) -- [fermion] (c),
      (m) -- [fermion] (d),
      };
    \end{feynman}
  \end{tikzpicture}
  \nonumber\\
 &  g^H_{4,2} + 2 g^H_{4,0} = -\frac{3}{2}\sqrt{c_s}^{3} (g^H_{3,3}+ g^H_{3,1})\\
 & \begin{tikzpicture}[baseline=(m.base)]
    \begin{feynman}
      \vertex[draw,circle,fill=black!20,minimum size=0.05cm] (m) at ( 0, 0) ;
    \vertex[draw,circle,fill=black!20,minimum size=0.05cm] (n) at ( 1, 0) ;
      \vertex (a) at (-1,-1) {};
      \vertex (b) at ( -1, 1) {};
      \vertex (c) at (2, 1) {};
      \vertex (d) at ( 2, -1) {};
      \vertex (e) at ( -1, 0) {};
      \diagram* {
      (m) -- [fermion] (a),
      (m) -- [fermion] (b),
      (m) -- [red, fermion] (e),
      (m) -- (n),
      (n) -- [fermion] (c),
      (n) -- [fermion] (d),
      };
    \end{feynman}
  \end{tikzpicture} 
  + 
  \begin{tikzpicture}[baseline=(m.base)]
    \begin{feynman}
      \vertex (m)[draw,circle,fill=black!20,minimum size=0.05cm] at ( 0, 0) ;
      \vertex(n)[draw,circle,fill=black!20,minimum size=0.05cm] at ( 1, 0) ;
      \vertex (a) at (-1,-1) {};
      \vertex (b) at ( -1, 1) {};
      \vertex (c) at (2, 1) {};
      \vertex (d) at ( 2, -1) {};
      \vertex (e) at ( 2, 0) {};
      \diagram* {
      (m) -- [fermion] (a),
      (m) -- [fermion] (b),
      (n) -- [red, fermion] (e),
      (m) -- (n),
      (n) -- [fermion] (c),
      (n) -- [fermion] (d),
      };
    \end{feynman}
  \end{tikzpicture}
  \rightarrow
    \begin{tikzpicture}[baseline=(m.base)]
    \begin{feynman}
      \vertex[draw,circle,fill=black!20,minimum size=0.05cm] (m) at ( 0, 0) ;
    \vertex[draw,circle,fill=black!20,minimum size=0.05cm] (n) at ( 1, 0) ;
      \vertex (a) at (-1,-1) {};
      \vertex (b) at ( -1, 1) {};
      \vertex (c) at (2, 1) {};
      \vertex (d) at ( 2, -1) {};
      \vertex (e) at ( 2, 0) {};
      \diagram* {
      (m) -- [fermion] (a),
      (m) -- [fermion] (b),
      (m) -- (n),
      (n) -- [fermion] (c),
      (n) -- [fermion] (d),
      };
    \end{feynman}
  \end{tikzpicture}\nonumber\\
&   g^H_{4,0} = -\frac{1}{4}\sqrt{c_s}^3 g^H_{3,1}\\
&  \begin{tikzpicture}[baseline=(m.base)]
    \begin{feynman}
      \vertex (m)[draw,circle,fill=black!20,minimum size=0.05cm] at ( 0, 0) ;
      \vertex (a) at (-1,-1) {};
      \vertex (b) at ( -1, 1) {};
      \vertex (c) at (1, 1) {};
      \vertex (d) at ( 1, -1) {};
      \vertex (e) at ( 1, 0) {};
      \diagram* {
      (m) -- [fermion] (a),
      (m) -- [fermion] (b),
      (m) -- [fermion] (c),
      (m) -- [fermion] (d),
      (m) -- [red,fermion] (e),
      };
    \end{feynman}
  \end{tikzpicture}
  +
  \begin{tikzpicture}[baseline=(m.base)]
    \begin{feynman}
      \vertex[draw,circle,fill=black!20,minimum size=0.05cm] (m) at ( 0, 0) ;
    \vertex[draw,circle,fill=black!20,minimum size=0.05cm] (n) at ( 1, 0) ;
      \vertex (a) at (-1,-1) {};
      \vertex (b) at ( -1, 1) {};
      \vertex (c) at (2, 1) {};
      \vertex (d) at ( 2, -1) {};
      \vertex (e) at ( 2, 0) {};
      \diagram* {
      (m) -- [fermion] (a),
      (m) -- [fermion] (b),
      (m) -- [dashed](n),
      (n) -- [fermion] (c),
      (n) -- [fermion] (d),
      (n) -- [red, fermion] (e),
      };
    \end{feynman}
  \end{tikzpicture} 
  + 
   \begin{tikzpicture}[baseline=(m.base)]
    \begin{feynman}
      \vertex (m)[draw,circle,fill=black!20,minimum size=0.05cm] at ( 0, 0) ;
      \vertex(n) at ( 1, 0) ;
      \vertex (a) at (-1,-1) {};
      \vertex (b) at ( -0.7, 0.7) {};
      \vertex (c) at (2, 1) {};
      \vertex (d) at ( 1.707, -0.707) ;
      \vertex (e) at ( 3.12, -0.707) {};
      \vertex (f) at ( 1, -2) {};
      \diagram* {
      (m) -- [fermion] (a),
      (m) -- [red, fermion] (b),
      (m) -- [dashed] (n),
      (n) -- [fermion] (c),
      (n) -- [dashed] (d),
      (d) -- [fermion] (e),
      (d) -- [fermion] (f),
      };
    \end{feynman}
  \end{tikzpicture}
  +
  \text{perms}
  \rightarrow \nonumber\\
&\begin{tikzpicture}[baseline=(m.base)]
    \begin{feynman}
      \vertex (m)[draw,circle,fill=black!20,minimum size=0.05cm] at ( 0, 0) ;
      \vertex (a) at (-1,-1) {};
      \vertex (b) at ( -1, 1) {};
      \vertex (c) at (1, 1) {};
      \vertex (d) at ( 1, -1) {};
      \vertex (e) at ( 1, 0) {};
      \diagram* {
      (m) -- [fermion] (a),
      (m) -- [fermion] (b),
      (m) -- [fermion] (c),
      (m) -- [fermion] (d),
      };
    \end{feynman}
  \end{tikzpicture}
  +
  \begin{tikzpicture}[baseline=(m.base)]
    \begin{feynman}
      \vertex (m)[draw,circle,fill=black!20,minimum size=0.05cm] at ( 0, 0) ;
      \vertex(n) at ( 1, 0) ;
      \vertex (a) at (-1,-1) {};
      \vertex (b) at ( -1, 1) {};
      \vertex (c) at (2, 1) {};
      \vertex (d) at ( 2, -1) {};
      \vertex (e) at ( 2, 0) {};
      \diagram* {
      (m) -- [fermion] (a),
      (m) -- [fermion] (b),
      (m) -- [dashed] (n),
      (n) -- [fermion] (c),
      (n) -- [fermion] (d),
      };
    \end{feynman}
  \end{tikzpicture}
  +
  \text{perms}
  \nonumber \\
 & g^H_{5,3} - 12 g^H_{3,3}g^H_{4,0} = \sqrt{c_s}^3(-g^H_{4,2} - 2g^H_{4,4}) ,\\
  &  g^H_{5,1} - 2 g^H_{3,1}g^H_{4,0} =\frac{1}{2}\sqrt{c_s}^3 (-g^H_{4,2} - 2g^H_{4,0}).
\end{align}
    \caption{Non-linear boost soft theorem constraints on superfluid Wilson coefficients. Black legs denote external hard momenta, red legs denote soft momenta, dashed internal legs represent the propagators that are effectively canceled by the vertices yielding terms regular in co-linear limit. The first line shows at level $n=3$ i.e. 4-point-3-point, there's one constraint on a linear combination of $g_{4,2}$ and $g_{4,0}$. At level $n=4$, we have three additional constraints, the second line shows 5-point exchange diagrams yield one constraint on $g_{4,0}$. The third line shows two 5-point constraints can be obtained by matching the regular terms at $n=4$.} 
    \label{SFcheck}
\end{figure}

\section{Scaling superfluid}\label{Sec6}

One particular type of superfluid is the Scaling Superfluid, in which the original theory possesses not only the full Poincaré symmetry but also an additional dilatation symmetry. The Scaling Superfluid action is given by \cite{Pajer:2018egx,Grall:2020ibl}
\begin{align}
    S = \int d^D x (\partial_\mu \phi \partial^\mu \phi)^\alpha.
\end{align}
where $\Delta$ is the scaling dimension of the scalar field and $\alpha = \frac{D}{2(\Delta + 1)}$. The theory for perturbation under the single clock background $\braket{\phi} = t$ takes the form of
\begin{align}
    \mathcal{L} = & [1 + 2\dot\pi + \dot\pi^2 - (\partial_i\pi)^2]^\alpha.
\end{align}
and $\alpha$ can be substituted by the 'sound speed' via $\alpha = \frac{c_s^2 + 1}{2c_s^2}$. Notably, the theory only allows for a perturbative $S$-matrix on the Lorentz breaking vacuum, since it is strongly coupled around the Lorentz invariant vacuum. It’s possible to extend the dilatation symmetry to full conformal symmetry, resulting in a Conformal Superfluid with a fixed 'sound speed' parameter. The algebraic classification and coset construction of scaling/conformal superfluid along with its derivation can be found in \cite{Pajer:2018egx, Green:2020ebl,Monin:2016jmo,Creminelli:2022onn}. Every Lagrangian coupling is fixed in terms of 'sound speed' parameter by
 \begin{align}
    &g^s_{3,3} = \frac{1}{6}(1 - c_s^2)\sqrt{\frac{c_s}{1+c_s^2}},\;\;g^s_{3,1} = -\frac{1}{2}(1-c_s^2)\sqrt{\frac{c_s}{1+c_s^2}},\nonumber\\
    &g^s_{4,4} = \frac{1}{24}\frac{c_s(1-c_s^2)(1 - 2c_s^2)}{(1+c_s^2)},\;\;g^s_{4,2} = \frac{1}{4}\frac{(1-c_s^2)(1 - 2 c_s^2)}{(1+c_s^2)},\nonumber\\
    &g^s_{4,0} = \frac{1}{8}\frac{(1-c_s^2)c_s}{(1+c_s^2)},\nonumber\\
    &g^s_{5,5} = \frac{1}{120}(1-c_s^2)(1 - 2c_s^2) (1 - 3 c_s^2) \sqrt{\frac{c_s}{1+c_s^2}}^3\nonumber\\
    &g^s_{5,3} = -\frac{1}{12}(1-c_s^2)(1 - 2c_s^2) (1 - 3 c_s^2)\sqrt{\frac{c_s}{1+c_s^2}}^3,\;\;g^s_{5,1} = \frac{1}{8}c_s^2(1-c_s^2)(1-3c_s^2)\sqrt{\frac{c_s}{1+c_s^2}}^3.\label{ScalingSuperfluidL}
\end{align}
The Hamiltonian Wilson coefficients are given by
 \begin{align}
    &g^{sH}_{3,3} = -\frac{1}{6}(1 - c_s^2)\sqrt{\frac{c_s}{1+c_s^2}},\;\;g^{sH}_{3,1} = \frac{1}{2}(1-c_s^2)\sqrt{\frac{c_s}{1+c_s^2}},\nonumber\\
    &g^{sH}_{4,4} = \frac{1}{24}\frac{c_s(1-c_s^2)(2-c_s^2)}{(1+c_s^2)},\;\;g^{sH}_{4,2} = -\frac{1}{4}\frac{c_s^3(1-c_s^2)}{(1+c_s^2)},\nonumber\\
    &g^{sH}_{4,0} = -\frac{1}{8}\frac{c_s^3(1-c_s^2)}{(1+c_s^2)},\nonumber\\
    &g^{sH}_{5,5} = -\frac{1}{120}(1-c_s^2)(6-5c_s^2+c_s^4)\sqrt{\frac{c_s}{1+c_s^2}}^3\nonumber\\
    &g^{sH}_{5,3} = \frac{1}{12}c_s^3(1-c_s^2)\sqrt{\frac{c_s}{1+c_s^2}},\;\;g^{sH}_{5,1} = -\frac{1}{8}c_s^2(1-c_s^2)(1-3c_s^2)\sqrt{\frac{c_s}{1+c_s^2}}^3.\label{ScalingSuperfluidH}
\end{align}
 Besides non-linear boost, the theory also non-linearly realises the dilatation symmetry from the original theory
 \begin{align}
     \delta_D\phi = - (\Delta + x^\mu\partial_\mu)\phi,
 \end{align}
by the perturbation around the background $\pi = \phi - t$,
\begin{align}
    \delta_D \pi = -t(\Delta + 1) - (\Delta + x^\mu \partial_{\mu})\pi. 
    \label{DilitationSym}
\end{align}
Without diving into a detailed derivation, we perform a coordinate and field rescaling $t\rightarrow \sqrt{2\alpha-1} t,\;\pi\rightarrow\pi/\sqrt{2\alpha\sqrt{(2\alpha-1)}}$ to canonically normalise the theory, the corresponding non-linear dilatation together with non-linear boost soft theorem are then
\begin{keyeqn}
\begin{align}
    \lim_{\vec{p}\rightarrow 0} \biggl\{ \frac{\tilde{\mathcal{A}}^{n+1}_{E_{p}}-\tilde{\mathcal{A}}^{n+1}_{-E_{p}}}{2E_{p}}  \biggr\} & = -i\sqrt{\frac{c_s^5}{1+c_s^2}}\biggl\{-2(n-2)\biggl(\frac{1-c_s^2}{c_s^2}\biggr)+\frac{1}{\Delta + 1}\bigl[n(-\Delta + D - 2) - D + 
 \vec{p}_a\cdot \partial_{\vec{p}_a}\bigr]\biggr\}  \tilde{\mathcal{A}}^n \,,\label{scale}\\
\lim_{\vec{p}\rightarrow{}0}  \biggl\{ \partial_{p^i} \left [\frac{\tilde{\mathcal{A}}^{n+1}_{E_{p}} + \tilde{\mathcal{A}}^{n+1}_{-E_{p}}}{2}\right]   \biggr\}&= -i \sqrt{\frac{c_s}{1+c_s^2}}\left[\sum_a E_a\partial_{p_a^i}\tilde{\mathcal{A}}^n\right] \,,\\
\lim_{\vec{p}\rightarrow 0} \biggl\{\tilde{\mathcal{A}}^{n+1}_{E_{p}} + \tilde{\mathcal{A}}^{n+1}_{-E_{p}} \biggr\} & =0\,.
\end{align}
\end{keyeqn}
The RHS of \eqref{scale} is reminiscent of the dilatation soft theorem derived in \cite{DiVecchia:2015jaq, Cheung:2021yog}. The amplitudes satisfy the non-linear boost soft theorem and the extra dilatation symmetry imposes additional constraints on the amplitudes. Now let's try to use the above soft theorem to bootstrap the Wilson coefficients. For $n = 3$, consider the 4-point and 3-point amplitudes respectively
\begin{align}
    &\tilde{\mathcal{A}}^3 = -i6(g^s_{3,3} + g^s_{3,1})E_2 E_3 (-E_2 - E_3),\\
    &\tilde{\mathcal{A}}^4_{E_1} = 24 [g^s_{4,4} - 2 (g^s_{3,1})^2 + 2 g^s_{3,1}(g^s_{3,3} + g^s_{3,1})] E_1 E_2 E_3 (- E_1 - E_2 - E_3) \nonumber\\
    &+ [g^s_{4,2} + 2 (g^s_{3,1})^2] \sum_{\text{perms}} E_1 E_2 (\vec{p}_3\cdot \vec{p}_4)  + \sum_{\text{perms}} g^s_{4,0} (\vec{p}_1\cdot\vec{p}_2)(\vec{p}_3\cdot\vec{p}_4)
    \;\;+\tilde{\mathcal{A}}^4|_{\text{pole}}.
\end{align}
The LHS \eqref{scale} yields
\begin{align}
    \lim_{\vec{p}_1\rightarrow 0} \biggl\{ \frac{\tilde{\mathcal{A}}^{4}_{E_{1}}-\tilde{\mathcal{A}}^{4}_{-E_{1}}}{2E_{1}}  \biggr\} = -12 \bigl[2g^s_{4,4} + g^s_{4,2} - 2 (g^s_{3,1})^2 + 4g^s_{3,1}(g^s_{3,1} + g^s_{3,3}) \bigr]E_2 E_3 (E_2 + E_3),
\end{align}
and the RHS yields
\begin{align}
    &-i\sqrt{\frac{c_s^5}{1+c_s^2}}\biggl\{-2\biggl(\frac{1-c_s^2}{c_s^2}\biggr) + \frac{1}{\Delta + 1}\bigl[3(-\Delta + D - 2) - D +\Vec{p}_2\cdot\partial_{\vec{p}_2} + \vec{p}_3\cdot\partial_{\vec{p}_3} \bigr]\nonumber\\
    \;\;&\bigl[-i6(g^s_{3,3} + g^s_{3,1})E_2 E_3 (-E_2 - E_3) \bigr]\biggr\}\\
    =&6(g^s_{3,3} + g^s_{3,1})\sqrt{\frac{c_s^5}{1+c_s^2}}\biggl[ -2\biggl(\frac{1-c_s^2}{c_s^2}\biggr) + \frac{1}{\Delta + 1}( -3\Delta + 2D - 3)\biggr]E_2 E_3 (E_2 + E_3).
\end{align}
Equating both sides we get
\begin{align}
    2g^s_{4,4} + g^s_{4,2}- 2 (g^s_{3,1})^2 + 6g^s_{3,1}(g^s_{3,1} + g^s_{3,3})= -\sqrt{\frac{c_s^5}{1+c_s^2}}\frac{-3\Delta + 2D - 3}{2(\Delta+1)}(g^s_{3,3} + g^s_{3,1}).
\end{align}
The $\Delta,\;D$ dependence can be re-expressed via $c_s$ by
\begin{align}
    \frac{ -3\Delta + 2D - 3}{\Delta+1} =\frac{2}{c_s^2} - 1.
\end{align}
We then arrive at
\begin{align}
    2g^s_{4,4} + g^s_{4,2}- 2 (g^s_{3,1})^2 + 6g^s_{3,1}(g^s_{3,1} + g^s_{3,3}) = -\frac{2 - c_s^2}{2}\sqrt{\frac{c_s}{1+c_s^2}}(g^s_{3,3} + g^s_{3,1}).
\end{align}
Once again the constraint is more manifest in the Hamiltonian basis
\begin{align}
    2g^{sH}_{4,4} + g^{sH}_{4,2} = -\frac{3\Delta - D + 3}{2(\Delta+1)}\sqrt{\frac{c_s^5}{1+c_s^2}} (g^{sH}_{3,3} + g^{sH}_{3,1})= \frac{(1-2c_s^2)}{2}\sqrt{\frac{c_s}{1+c_s^2}}(g^{sH}_{3,3} + g^{sH}_{3,1}).
\end{align}
Notice $g_{4,0}^s$ is unconstrained by the non-linear dilatation soft theorem alone, however since the original theory preserves boost and from the last section we know the non-linear boost soft theorem would give us two more constraints from $n=3\; \text{and}\; 4$
\begin{align}
    g^{sH}_{4,2} + 2 g^{sH}_{4,0} = -\frac{3}{2} \sqrt{\frac{c_s^5}{1+c_s^2}} (g^{sH}_{3,3} + g^{sH}_{3,1}),\;
    g^{sH}_{4,0} = -\frac{1}{4}\sqrt{\frac{c_s^5}{1+c_s^2}} g^{sH}_{3,1}.
\end{align}
therefore the soft theorems uniquely fix all the 4-point parameters given 3-point parameters. The analysis for five point is identical to the case of non-linear boost, the extra constraint from the non-linear dilatation symmetry at 5-point is 
\begin{align}
    5g^{sH}_{5,5} - 24g^{sH}_{3,3}g^{sH}_{4,4} =  \frac{4\Delta - D + 4}{\Delta+1}\sqrt{\frac{c_s^5}{1+c_s^2}} g^{sH}_{4,4} = c_s^3\frac{3c_s^2 - 1}{1+c_s^2}g^{sH}_{4,4}.
\end{align}
Together with non-linear boost constraints we're able to bootstrap all 4-point and 5-point coefficients given the 3-point coefficients and one is able to check the result agrees with the Hamiltonian. For higher point bootstrap, we get one extra constraint at each point relative to the non-linear boost case, thus the remaining DoF $\dot\pi^n$ that is unconstrained at each order in $\pi$ by non-linearly realised boost is fixed by the non-linearly realised dilatation. 

However, that is still not completely satisfying, since the 3-point amplitude is still undetermined. One might be curious if we're able to use soft theorems to determine the 3-point coefficients. The answer is yes if we tweak the two point function. In order to get constraints from $n=2$, we need to abandon the all-out formalism as at $n=2$ the energy and momentum delta function are not independent. The two point function at tree level for $\pi(p_2)\rightarrow\pi(p_3)$ is given by
\begin{align}
    \braket{\vec{p}_2|\vec{p}_3} = 2 E_2 \delta(0) \delta^{D-1}(\vec{p}_2-\vec{p}_3),
\end{align}
where $\delta(0)$ is a dimensionless normalisation factor.
The RHS of the soft theorem at $n=2$ is then
\begin{align}
    &-i\sqrt{\frac{c_s^5}{1+c_s^2}}\biggl[  \frac{1}{\Delta + 1}(2\Delta - 2D + 4 - \sum_a \vec{p}_a\cdot\partial_{\vec{p}_a})\biggr] 2 E_2 \delta(0) \delta^{D-1}(\vec{p}_2-\vec{p}_3)\nonumber\\
    =& -i\frac{1}{\Delta + 1}\sqrt{\frac{c_s^5}{1+c_s^2}} (2\Delta - D + 2) 2 E_2 \delta(0) \delta^{D-1}(\vec{p}_2-\vec{p}_3).
\end{align}
The LHS is
\begin{align}
    \lim_{\vec{p}_1\rightarrow 0}\frac{\tilde{\mathcal{A}}^3_{E_1} - \tilde{\mathcal{A}}^3_{-E_1}}{2E_1} &= i6(g^s_{3,3} + g^s_{3,1}) E_2^2 \delta(E_2-E_3)\delta^{D-1}(\vec{p}_2-\vec{p}_3) \nonumber\\
    &=i6(g^s_{3,3} + g^s_{3,1}) E_2 \delta(0) \delta^{D-1}(\vec{p}_2-\vec{p}_3),
\end{align}
where the last equal sign is achieved by $\delta(E_2- E_3) = \frac{1}{E_2}\delta(1-\frac{E_2}{E_3}) \rightarrow \frac{1}{E_2}\delta(0)$. The constraints on the coefficients are then
\begin{align}
    6(g^s_{3,3} + g^s_{3,1}) = -2\sqrt{\frac{c_s^5}{1+c_s^2}} \frac{2\Delta - D + 2}{\Delta+1} = 2(1-c_s^2)\sqrt{\frac{c_s^5}{1+c_s^2}}.
\end{align}
The result again matches the explicit Lagrangian Wilson coefficients. The 3-point amplitude is completely fixed by the soft theorem. It is rather unexpected that the soft theorem could fix 3-point coefficient given a strong enough symmetry as the two point amplitude is usually considered to be zero. However, as illustrated by the scaling superfluid example, the full $S$-matrix with energy momentum delta function included is indispensable for the soft theorem to function properly, especially for the connection between 2 and 3-point. Therefore, we're able to conclude the soft theorems fix all the amplitude coefficients for the scaling superfluid up to the 'sound speed' $c_s$, we've explicitly checked up to 5-point amplitude.

\section{Hidden enhanced Adler zeros and field basis independence} \label{Sec7}
In the above example, we only considered diagrams that do not involve a soft cubic vertex that generates poles located at $s_{1,a} = 0$. This is expected since the kinematic dependence for such diagrams is essentially different from other topologies, for instance the analytical structure is disparate, therefore the collection of 4-point exchange diagrams that are singular in $s_{1,a} = 0$ on the LHS of the soft theorem have to vanish independently i.e.
\begin{align}
    \lim_{\vec{p}_1\rightarrow 0} \biggl\{ \frac{\tilde{\mathcal{A}}^{4}_{E_{1}}-\tilde{\mathcal{A}}^{4}_{-E_{1}}}{2E_{1}}|_{\text{pole}}  \biggr\} &= 0,\label{Exchange1}\\ 
        \lim_{\vec{p}_1\rightarrow 0} \biggl\{ \partial_{\vec{p}_1}\biggl[\frac{\tilde{\mathcal{A}}^{4}_{E_{1}} + \tilde{\mathcal{A}}^{4}_{-E_{1}}}{2}\biggr]|_{\text{pole}}  \biggr\} &= 0.\label{Exchange2}
\end{align}
This is not true at first glance, since the 3-point amplitude itself does not satisfy the soft theorem for scaling superfluid, i.e. it does not enjoy the enhanced soft scaling $\mathcal{O}(p_1^2)$. Therefore, non-trivial cancellations are required for the soft theorem to hold. There is indeed an enhanced soft scaling when the soft hierarchy is considered and after summing over all three channels, resulting in a surprising cancellation, the soft amplitude scales as
\begin{align}
    \tilde{\mathcal{A}}^4_{E_1}|_{\text{pole}} &= g^2\frac{(E_1+E_2)(E_1 E_2)(E_3 + E_4)(E_3 E_4) }{2(s_{1,a} + i\varepsilon E_{12})}+ 2\;\text{perms} \nonumber \\
    &= \mathcal{O}(p_1^2).
\end{align}
where $g$ is the cubic Wilson coefficient of the 3-point amplitude $\mathcal{A}_3 = g E_1 E_2 (E_1 + E_2)$.
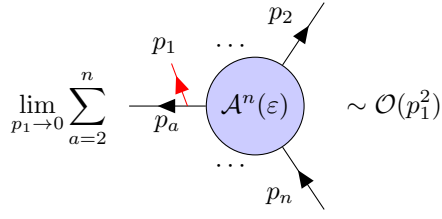
\begin{figure}
    \centering
\[ \lim_{p_1\rightarrow 0}\sum_{a=2}^n
\begin{tikzpicture}[baseline=(m.base)]
    \begin{feynman}
      \vertex[draw,circle,fill=blue!20,minimum size=1.3cm] (m)  at ( 0, 0) {$\mathcal{A}^{n}(\varepsilon)$}{};
      \vertex (b) at ( 1,-1.5) {};
      \vertex (c) at (-1.8, 0) {};
      \vertex (d) at ( 1, 1.5) {};
      \vertex (e) at ( -0.3, 0.8) {$\cdots$};
      \vertex (f) at ( -0.3, -0.8) {$\cdots$};
      \vertex (g) at ( -1.2, 0.8) {$p_1$};
      \vertex (h) at (-0.9,0) ;
      \diagram* {
        (m) -- [fermion, edge label = $p_a$] (c),
        (b) -- [fermion,  edge label = $p_n$] (m) -- [fermion, edge label=$p_{2}$] (d),
        (h) -- [fermion, red] (g),
      };
    \end{feynman}
  \end{tikzpicture} 
   \sim \mathcal{O}(p_1^2)
\]
    \caption{The sum of diagrams that contain 3-point soft vertices have an enhanced Adler zero $\mathcal{O}(p_1^2)$. Here we impose soft hierarchy by fixing $\varepsilon$ and send the soft momentum $p_1$ to zero first.} 
    \label{HiddenAdlerzero}
\end{figure}
This suggests higher point amplitudes that factorise into 3-point soft amplitudes would vanish independently on the LHS of the soft theorems. In fact this is quite easy to show for any point, the contributions to the scattering amplitude from the collection of diagrams where the soft momentum is attached to a cubic vertex and becomes singular in the co-linear limit $s_{1,a} \rightarrow 0$ are
\begin{align}
    \tilde{\mathcal{A}}_{E_1}^{n+1} \supset& \sum_{a=2}^{n+1}g\frac{(E_1+E_a)E_1 E_a}{2(s_{1,a} + i\varepsilon E_{1a})}\biggl( 1 + \vec{p_1}\cdot \partial_{\vec{p}_a} \biggr)\tilde{\mathcal{A}}^n\nonumber\\
    =&\sum_{a=2}^{n+1}\frac{g E_{1} E_{a}}{2i\varepsilon}\tilde{\mathcal{A}}^n + \mathcal{O}(p_1^2)\nonumber\\
    =& -\frac{g E_1^2}{2i\varepsilon}\tilde{\mathcal{A}}^n + \mathcal{O}(p_1^2)\nonumber\\
    =&\mathcal{O}(p_1^2).
\end{align}
To arrive from the second to third line we imposed energy-momentum conservation. A diagrammatic representation is shown in Figure \ref{HiddenAdlerzero}. Therefore for both the superfluid and the scaling superfluid scenarios, the collection of exchange diagrams whose soft momentum is attached to a cubic vertex that are singular in $s_{1,a} = 0$ doesn't contribute non-trivially to the soft theorem i.e. the LHS does not cancel with RHS. The cubic independence agrees with the soft theorem for unequal time correlators \cite{Hui:2022dnm}. It additionally offers a new cancellation mechanism of the explicit off-shell cubic vertex dependence in the scattering amplitude soft theorem in \cite{Green:2022slj,Cheung:2023qwn}. Moreover, we also see that the $i\varepsilon$ prescription plays a major role for the vanishing identity, i.e. the LHS of soft theorems for scaling superfluid \eqref{scale} does not vanish unless we choose our propagator to be
\begin{align}
    \Delta(p_a + p_b) = \frac{1}{2(s_{a,b} + i\varepsilon E_{ab})}.
\end{align}
The soft theorem would fail to hold if we absorb the energy dependence of the imaginary shift into $\varepsilon$
\begin{align}
    \Delta(p_a + p_b) = \frac{1}{2(s_{a,b} + i\varepsilon)}.
\end{align} 
Additionally, the soft theorems are field redefinition invariant under the minimal basis if the $i\varepsilon$ prescription is applied solely at co-linear poles located at 
$s_{1,a} = 0$ since the poles are indeterminate under the soft limit, serving as an IR regulator for the phase space integral. This again implies we should only make the $i\varepsilon$ shift on terms that are indeterminate or divergent in the soft limit. Specifically, we treat
\begin{align}
    \frac{s_{1,a}}{s_{1,a} + i\varepsilon E_{1,a}} \rightarrow 1.
\end{align}
which agrees with the derivation in Sec \ref{Sec2}. It is worth noting that, under the minimal basis, such $i\varepsilon$ shift is uniquely defined. The prescription ensures the field basis independence since both sides of the soft theorems only depend on on-shell data, and the conclusion aligns with the findings of \cite{Green:2022slj, Cheung:2023qwn}.

\section{Conclusion and outlook}
While using soft limits to classify and constrain Lorentz-invariant scalar EFTs is well understood, this approach is less developed for boost-breaking cases, which are relevant in condensed matter systems and cosmology. The scattering amplitude soft theorems are plagued by the presence of the cubic vertices, since one has to explicitly subtract the factorised contribution from soft cubic vertices. In this work, we aim at eliminating the off-shell vertex dependence by introducing a new cancellation mechanism by imposing the soft hierarchy.

We derived scattering amplitude soft theorems for generic non-linear symmetries and demonstrated that they are model-independent, as the soft theorems do not depend on unconstrained off-shell cubic vertices. During the derivation, we introduced a new order of limits, specifically using $i\varepsilon$ as an IR regulator for both soft and on-shell limits, which is essential for the all-order-perturbative validity of the soft theorems. In the cases of superfluid and scaling superfluid, we showed that the soft theorems are powerful enough to bootstrap all Wilson coefficients up to the known degrees of freedom given 'sound speed' parameter $c_s$. The result aligns with the findings from Hamiltonian analysis. For scaling superfluid, even the 3-point amplitude can be bootstrapped, thus the theory can be uniquely determined up to the 'sound speed' $c_s$. However, due to the degeneracy of 3-point on-shell amplitude, we're unable to determine the constraint for $g_{3,1}$ from amplitude soft theorem. The hidden enhanced Adler zero accounts for the off-shell cubic vertices dependence of the soft theorems for superfluid derived in \cite{Green:2022slj, Cheung:2023qwn}. Rather than explicitly subtracting the diagrammatic contribution from the soft cubic vertices, the soft hierarchy imposes an enhanced soft scaling across these diagrams collectively. Such prescription is essential for the soft amplitude combination $\mathcal{A}_{E_p} + \mathcal{A}_{-E_p}$ to be consistent with cancellation against any hard process. Finally, we commented that field basis independence is guaranteed by the $i\varepsilon$ shift exclusively on terms that are indeterminate or divergent under the limits being considered. Terms that are regular and determinate in the soft limit remain unaffected by the  $i\varepsilon$ shift.
There are few possible future avenues worth exploring:
\begin{itemize}
    \item In the present work, we restricted the inflaton perturbation to linear dispersion relations $E = |\vec{p}|$, as we take the free theory
    \begin{align}
       \mathcal{L}_2 = \frac{1}{2}[\dot\pi^2 - (\partial_i\pi)^2]
    \end{align}
    as an input. However, in the context of EFT of inflation \cite{Cheung:2007st}, contribution from $(\partial_i\pi)^2$ may be suppressed so that the leading contribution comes from $(\partial_i^2\pi)^2$, with ghost condensate \cite{Arkani-Hamed:2003pdi} serving as an example. It is thus natural to extend the scope of the current work to non-linear dispersion relations.
    \item We derived soft theorems for superfluid and scaling superfluid and showed they fix all Wilson coefficients up to know degrees of freedom given $c_s$. In the case of scaling superfluid, if $\Delta = 0$, then the dilatation symmetry is uplifted to full conformal symmetry, the free parameter $c_s$ in scaling superfluid is then fixed by the additional special conformal symmetry \cite{Pajer:2018egx}. It would be interesting to derive soft theorem for conformal superfluid and show it further fixes the 'sound speed' $c_s$.
    \item The ultimate goal of this program is to bootstrap the cosmological correlators in the context of EFToI with broken de-Sitter boost, thus it would be interesting to derive soft theorems or consistency relations for wavefunction coefficients \cite{Bittermann:2022nfh} and correlators\cite{ Creminelli:2012ed, Hui:2022dnm, Qin:2024gtr}, and then compare them to the soft theorems for scattering amplitudes by taking the residue of the $k_T$ pole \cite{Maldacena:2011nz, Pajer:2020wnj}. Furthermore, the constraints from the scattering amplitude soft theorems are evidently more structured in terms of Hamiltonian coefficients, which suggest that one could directly work at the level of correlators. 
    \item In \cite{Grall:2020ibl,Green:2022slj}, it was argued scalar DBI (Dirac-Born-Infeld) is the unique $P(X)$ theory that recovers Lorentz invariance under the boost-breaking background $\braket{\phi} = t$. It would be nice to apply the soft theorem derived in this work to achieve the same conclusion.
\end{itemize}

\paragraph{Acknowledgements} The author thanks Nima Arkani-Hamed, Oliver Gould, Austin Joyce, Scott Melville, Bartosz Pyszkowski, David Stefanyszyn and Takuya Yodafor their insightful discussions. Special thanks to David Stefanyszyn for his unwavering support and for meticulously reviewing the draft. ZD is supported by Nottingham CSC [file No. 202206340007].
For the purpose of open access, the authors have applied a CC BY public copyright licence to any Author
Accepted Manuscript version arising.

%%%%%%%%%%%%%%%%%%%%%%%%%%%%%%%%%%

%%%%%%%%%%%%%%%%%%%%%%%%%%%%%%%%%%%%%%%%%%%%%%%%%%%%%%%%%%%%%%%%%%%%%%%%%

\appendix
\section{More on Hamiltonian of superfluid EFT}\label{LagHam}
Schematically, the Lagrangian of a generic Lorentz symmetry violating scalar theory with one derivative per field after canonical normalisation truncating up to five point is
\begin{align}
    \mathcal{L} = &\frac{1}{2} [\dot\pi^2 - (\partial_i\pi)^2] + g_{3,3} \dot \pi^3 + g_{3,1} \dot\pi (\partial_i\pi)^2 \nonumber\\
    & + g_{4,4} \dot\pi^4 + g_{4,2} \dot\pi^2 (\partial_i\pi)^2 + g_{4,0} (\partial_i\pi)^4 \nonumber\\
    & + g_{5,5} \dot\pi^5 + g_{5,3} \dot\pi^3 (\partial_i\pi)^2 + g_{5,1} \dot\pi (\partial_i\pi)^4 \ldots
\end{align}
We denote the conjugate momentum as $\tilde{\pi}$, it is given by
\begin{align}
    \tilde{\pi} = \frac{\delta \mathcal{L}}{\delta\dot\pi} =& \dot\pi + 3 g_{3,3} \dot\pi^2 + g_{3,1} (\partial_i\pi)^2 \nonumber\\
    +& 4 g_{4,4}\dot\pi^3 + 2 g_{4,2} \dot\pi^2 (\partial_i\pi)^2 \nonumber\\
    +& 5 g_{5,5} \dot\pi^4 + 3 g_{5,3} \dot\pi^2 (\partial_i\pi)^2 + g_{5,1} (\partial_i\pi)^4 + \ldots
\end{align}
The Hamiltonian is thus
\begin{align}
    \mathcal{H}(\tilde{\pi}, \pi) &= \tilde\pi \dot\pi - \mathcal{L} \nonumber\\ 
    &= \frac{1}{2} [\tilde{\pi}^2 + (\partial_i\pi)^2] - g_{3,3} \tilde \pi^3 - g_{3,1} \tilde\pi (\partial_i\pi)^2 \nonumber\\
    &+ \frac{9}{2}[(g_{3,3})^2 - g_{4,4}] \tilde\pi^4 + [3g_{3,3}g_{3,1} - g_{4,2}] \tilde\pi^2(\partial_i\pi)^2 + [\frac{1}{2} (g_{3,1})^2 - g_{4,0}] (\partial_i\pi)^4 \nonumber\\
    &+ [-27 (g_{3,3})^3 + 12 g_{3,3} g_{4,4} - g_{5,5}] \tilde\pi^5 + [-18 (g_{3,3})^2 g_{3,1} + 6 g_{3,3} g_{4,2} + 4 g_{3,1}g_{4,4} - g_{5,3}] \tilde\pi^3 (\partial_i\pi)^2 \nonumber\\
    &+ [-3 g_{3,3} (g_{3,1})^2 + 2 g_{3,1} g_{4,2} - g_{5,1}] \tilde\pi(\partial_i\pi)^4 + \ldots \label{Ham}
\end{align}
In the context of superfluid, the Lagrangian could be parameterised as
\begin{align}
    \mathcal{L} = X + M_2 X^2 + M_3 X^3 + M_4 X^4 + M_5 X^5 + \ldots
\end{align}
where $X = \partial_\mu \phi \partial^\mu \phi - 1 = \dot\pi + \frac{1}{2}\dot\pi^2 - \frac{1}{2}(\partial_i\pi)^2 $. Upon canonical normalisation, the Lagrangian Wilson coefficients are
\begin{align}
    &g_{3,3} = \sqrt{c_s^{7}} \biggl(\frac{c_s^2-1}{2c_s^2} + M_3 \biggr),\;\;g_{3,1} =  \frac{1}{2\sqrt{c_s}}(c_s^2 - 1),\\
    &g_{4,4} = \frac{1}{8} c_s^3 [1 + c_s^2(8M_4 + 12 M_3 -1)],\;\;g_{4,2} = \frac{c_s}{4}[c_s^2(1 - 6M_3) - 1],\;\;g_{4,0} = \frac{1-c_s^2}{8c_s},\\
    &g_{5,5} = \frac{\sqrt{c_s^{13}}}{4} (3M_3 + 8M_4 + 4M_5),\;\;g_{5,3} = -\frac{\sqrt{c_s^9}}{2} (3M_3 + 4M_4),\;\;g_{5,1} = \frac{3\sqrt{c_s^5}}{4}M_3,
\end{align}
where we reparameterised $M_2$ by $M_2 = \frac{1}{c_s^2} - 1$. Applying the Hamiltonian-Lagrangian conversion from \eqref{Ham}, the Hamiltonian Wilson coefficients $g^H_{m,n}$ are
\begin{align}
    &g^H_{3,3} = -\sqrt{c_s^{7}} \biggl(\frac{c_s^2-1}{2c_s^2} + M_3\biggr),\;\;g^H_{3,1} = -\frac{1}{2\sqrt{c_s}}(c_s^2 - 1),\nonumber\\
    &g^H_{4,4} = \frac{1}{8} c_s^3 \biggl[8 + 9c_s^4(1 - 2M_3)^2 + c_s^2(-8M_4 + 24M_3 -17)\biggr],\;\;g^H_{4,2} = \frac{c_s}{4} \biggl[c_s^4(6M_3 - 3) + 5c_s^2 -2\biggr],\nonumber\\
    &g^H_{4,0} = \frac{c_s(c_s^2 -1)}{8},\nonumber\\
    &g^H_{5,5} = -\frac{\sqrt{c_s^9}}{8}\biggl\{21 + c_s^2\biggl[-69 + 84M_3 + 3c_s^2(2M_3 -1)(-25 + 9c_s^2(2M_3-1)^2 + 30M_3 -16M_4) - 32M_4 + 8M_5\biggr] \biggr\}\nonumber\\
    &g^H_{5,3} = \frac{c_s^3}{8}\biggl[8 + 9c_s^4(2M_3 - 1)^2 + c_s^2(-17 + 24M_3 -8M_4)\biggr],\;\;g^H_{5,1} = \frac{\sqrt{c_s}}{8}\biggl[-1 + 5c_s^2 -7 c_s^4 + c_s^6(3 - 6M_3)\biggr]
\end{align}
The computation is consistent with the results in \cite{Chen:2006dfn, Hui:2022dnm}

%%%%%%%%%%%%%%%%%%%%%%%%%%%%%%%%%%%%%%%%%%%%%%%%%%%%

\bibliographystyle{JHEP}
\bibliography{refsAdlerZero}

\end{document}